# A Thermodynamic Theory of Proximity Ferroelectricity

Eugene A. Eliseev[1], Anna N. Morozovska[2*], Jon-Paul Maria[3], Long-Qing Chen[3†], and Venkatraman Gopalan[3‡]

[1] Frantsevich Institute for Problems in Materials Science, National Academy of Sciences of Ukraine, 3, str. Omeliana Pritsaka, 03142 Kyiv, Ukraine

[2] Institute of Physics, National Academy of Sciences of Ukraine, 41, pr. Nauki, 03028 Kyiv, Ukraine

[3] Department of Materials Science and Engineering, Pennsylvania State University, University Park, PA 16802, USA

**Abstract**

Proximity ferroelectricity has recently been reported as a new design paradigm for inducing ferroelectricity, where a non-ferroelectric polar material becomes a ferroelectric by interfacing with a thin ferroelectric layer. Strongly polar materials, such as AlN and ZnO, which were previously unswitchable with an external field below their dielectric breakdown fields, can now be switched with practical coercive fields when they are in intimate proximity to a switchable ferroelectric. Here, we develop a general Landau-Ginzburg theory of proximity ferroelectricity in multilayers of non-ferroelectrics and ferroelectrics to analyze their switchability and coercive fields. The theory predicts regimes of both "proximity switching" where the multilayers collectively switch, as well as "proximity suppression" where they collectively do not switch. The mechanism of the proximity ferroelectricity is an internal electric field determined by the polarization of the layers and their relative thickness in a self-consistent manner that renormalizes the double-well ferroelectric potential to lower the steepness of the switching barrier. Further reduction in the coercive field emerges from charged defects in the bulk that act as nucleation centers. The application of the theory to proximity ferroelectricity in $Al_{x-1}Sc_xN/AlN$ and $Zn_{1-x}Mg_xO/ZnO$ bilayers is demonstrated. The theory further predicts that multilayers of dielectric/ferroelectric and paraelectric/ferroelectric layers can potentially result in induced ferroelectricity in the dielectric or paraelectric layers, resulting in the entire stack being switched, an exciting avenue for new discoveries. This thawing of "frozen ferroelectrics", paraelectrics and potentially dielectrics with high dielectric constant, promises a large class of new ferroelectrics with exciting prospects for previously unrealizable domain-patterned optoelectronic and memory technologies.

---

[*] Corresponding author, e-mail: anna.n.morozovska@gmail.com

[†] Corresponding author, e-mail: lqc3@psu.edu

[‡] Corresponding author, e-mail: vgopalan@psu.edu

## 1. Introduction

The recently discovered lead-free wurtzite and fluorite ferroelectrics, such as nanoscale $Hf_xZr_{1-x}O_2$, $Zn_{1-x}Mg_xO$, $Al_{x-1}Sc_xN$ and $Al_{x-1}B_xN$, are promising candidates for the next-generation of Si-compatible electronic memory elements such as ferroelectric random-access memories (FeRAMs), steep-slope field-effect transistors (FETs), and other logic devices [1, 2, 3]. In all these examples, the atomic-scale chemical stress through bulk doping stabilizes a ferroelectric phase (as e.g., in Zr doping of $HfO_2$) or dramatically lowers the coercive field for ferroelectric switching to practical levels (as in Mg doping of polar but unswitchable ZnO, and Sc or B doping of the polar but unswitchable AlN) [4, 5, 6, 7]. The local lattice distortions and charged defects around chemical dopants can locally allow for polarization switching to create domain nuclei which then propagate through long range electrostatic forces [8, 9]. In particular, polarization switching in wurtzite and fluorite ferroelectrics with a large coercive field $E_c$ (such as $Hf_xZr_{1-x}O_2$, $Zn_{1-x}Mg_xO$ and $Al_{x-1}Sc_xN$, where $E_c > 1 - 5$ MV/cm) proceeds from "extrinsic" sources of domain nucleation such as charged or/and elastic defects, which can lead to the correlated nucleation of the spike-like domains in $Al_{x-1}Sc_xN$/GaN layers [4] and in $Zn_{1-x}Mg_xO$ layers [7]. Correlated nucleation of the spike-like domains [10], as well as correlated polarization switching in the proximity of ferroelectric domain walls [11], have also been revealed in ferroelectric $LiNbO_3$ (with $E_c > 1$ MV/cm). However, bulk doping can also have deleterious effects on functional properties, such as significant optical scattering losses that impede optoelectronic applications [12]. A question arises as to whether it is possible to apply such stress only at the surfaces of an otherwise non-switchable polar material to trigger ferroelectric switching without breakdown. In other words, can surface engineering by itself thaw "frozen ferroelectrics" such as AlN and ZnO without any need for bulk doping, leaving the bulk pristine but switchable?

The recent observation of "proximity ferroelectricity" by Skidmore et al. [13] experimentally confirms that indeed this is possible in wurtzite ferroelectric heterostructures. They define this phenomenon as follows: [quote] "Proximity ferroelectricity is an interface-associated phenomenon, where electric field driven polarization reversal in a non-ferroelectric polar material is induced by one or more adjacent ferroelectric materials". The proximity ferroelectricity was revealed experimentally in the non-ferroelectric layers (such as AlN and ZnO) coupled with the ferroelectric layers (such as $Al_{1-x}B_xN$, $Al_{1-x}Sc_xN$ and $Zn_{1-x}Mg_xO$) in nitride-nitride, oxide-oxide, and nitride-oxide multilayers. The layered structures, whose thicknesses varied from tens to hundreds of nm, included two-layer (asymmetric, e.g. $Al_{1-x}Sc_xN$/AlN, $Al_{1-x}B_xN$/AlN, ZnO/$Al_{1-x}B_xN$) and three-layer (symmetric, e.g. $Al_{1-x}B_xN$/AlN/$Al_{1-x}B_xN$, AlN/$Al_{1-x}B_xN$/AlN, $Zn_{1-x}Mg_xO$/ZnO/$Zn_{1-x}Mg_xO$) configurations [13]. These exciting experimental breakthroughs provide a pathway to discover entirely new classes of ferroelectrics with pristine materials without bulk chemical doping. But how and when does this proximity mechanism work, and what else is possible? Can a paraelectric or a dielectric layer, for example, also exhibit proximity ferroelectricity? It appears theoretically possible, as we demonstrate next.

While the Landau-Ginzburg-Devonshire (LGD) phenomenological description of materials such as $Al_{1-x}Sc_xN$ is available [14], to the best of our knowledge, a phenomenological description of the proximity effect is currently absent. In this work we consider multilayer films consisting of separate layers with different thicknesses, which have different ferroelectric, paraelectric and dielectric properties, in the framework of the LGD phenomenological approach. We show that either all of the different-types of layers undergo the simultaneous ferroelectric switching in a very similar manner (dubbed "proximity switching") [13], or all of the different-types of layers demonstrate paraelectric or non-switching behavior simultaneously due to the strong electric coupling via the internal fields (dubbed "proximity suppression"). The proposed mechanism of the bilayer switching is used to describe quantitatively the proximity-induced polarization switching of defect-free bulk layers in $Al_{x-1}Sc_xN/AlN$ bilayers from purely interfacial electrostatic considerations alone. Additional reduction of the coercive field is shown to be achieved from including random electric charges within the bulk, which act as nucleation centers for the correlated polarization switching.

## 2. Problem statement

Let us consider the multilayer film consisting of $N$ separate layers with thickness $h_i$ (with subscript "$i$" denoting different layers), which have different ferroelectric and dielectric properties. At least one of the layers is a uniaxial ferroelectric with polarization $\vec{P}^{(i)}$ directed along the polar axis Z-axis pointed along the multilayer normal vector $\vec{e}_z$. In addition, at least one of the layers is a paraelectric material (i.e. an "incipient" ferroelectric with a negative Curie temperature), or a dielectric material (without a soft phonon mode), or a non-switchable ("frozen") polar material with a coercive field larger than its dielectric breakdown field. In all these cases the free energy of the system is the sum of the Landau-Ginzburg Devonshire (LGD) bulk part $F_b$ and surface/interfacial energies $F_s$. The bulk part $F_b$ is

$$F_b = \sum_{i=1}^{N} \int_{V_i} \left[ \frac{\alpha_i}{2}\left(P_z^{(i)}\right)^2 + \frac{\beta_i}{4}\left(P_z^{(i)}\right)^4 + \frac{\gamma_i}{6}\left(P_z^{(i)}\right)^6 + \frac{g_z^{(i)}}{2}\left(\frac{\partial P_z^{(i)}}{\partial z}\right)^2 + \frac{g_\perp^{(i)}}{2}\left\{\left(\frac{\partial P_z^{(i)}}{\partial y}\right)^2 + \left(\frac{\partial P_z^{(i)}}{\partial x}\right)^2\right\} - P_z^{(i)}E_z^{(i)} - \frac{\varepsilon_0 \varepsilon_b^{(i)}}{2}\left(E_j^{(i)}\right)^2 \right] dV \quad (1a)$$

Here $\alpha_i$, $\beta_i$ and $\gamma_i$ are LGD expansion coefficients of the individual layers, respectively. $E_j^{(i)}$ is the $j$-th component of the electric field $\vec{E}^{(i)}$ inside the layer "$i$". The last term, $\frac{\varepsilon_0 \varepsilon_b^{(i)}}{2}\left(E_j^{(i)}\right)^2$, where the summation over $j$ is performed ($j = x, y$, or $z$), accounts for the possible appearance of transverse electric field via the 2D or 3D domain formation and/or random charge defects. Hereafter, $\varepsilon_b^{(i)}$ is the background dielectric permittivity [15, 16] of the layer "$i$". The value of $\varepsilon_b^{(i)}$ could potentially vary from 3 to 90, but typically does not exceed 10 [15]. The electric field $\vec{E}^{(i)}$ can be found from the Poisson-type equations for the electrostatic potential $\varphi$ in each layer with appropriate boundary conditions, which are the fixed potential at the electrodes, potential and electric displacement continuity at the interfaces between the layers. The interfacial energy contribution is

$$F_{int} = \sum_{i=1}^{N} \int_{S_{i,i+1}} \left(\frac{a_i}{2}\left(P_z^{(i)}\right)^2 + \frac{a_{i+1}}{2}\left(P_z^{(i+1)}\right)^2 - b_{i,i+1} P_z^{(i)} P_z^{(i+1)}\right) dS, \qquad (1b)$$

where $S_{i,i+1}$ is the surface area of the interface between $i$-th and $(i+1)$-th layers, and $a_i$, $a_{i+1}$ and $b_{i,i+1}$ are the coefficients of interface energy. Notably, the inequality $|b_{i,i+1}| \leq \sqrt{a_i a_{i+1}}$ should be valid for the surface energy stability. The surface energy corresponding to the layers "1" and "$N$" contacting with the electrodes "$e$" is

$$F_{sur} = \int_{S_{1,e}} \frac{a_1}{2}\left(P_z^{(1)}\right)^2 dS + \int_{S_{N,e}} \frac{a_N}{2}\left(P_z^{(N)}\right)^2 dS. \qquad (1c)$$

With the total free energy, $F_b + F_{int} + F_{sur}$, the time-dependent LGD (TLGD) equations for each layer and the interfacial boundary conditions are derived:

$$\Gamma_i \frac{\partial P_z^{(i)}}{\partial t} + \alpha_i P_z^{(i)} + \beta_i \left(P_z^{(i)}\right)^3 + \gamma_i \left(P_z^{(i)}\right)^5 - g_z^{(i)} \frac{\partial^2 P_z^{(i)}}{\partial z^2} - g_\perp^{(i)} \left(\frac{\partial^2 P_z^{(i)}}{\partial x^2} + \frac{\partial^2 P_z^{(i)}}{\partial y^2}\right) = E_z^{(i)}. \qquad (2a)$$

Here $\Gamma_i$ are the Landau-Khalatnikov relaxation coefficient of the layer "$i$". The boundary conditions at the $\{i, i+1\}$-interface are:

$$a_i P_z^{(i)} + g_z^{(i)} \frac{\partial P_z^{(i)}}{\partial z} = b_{i,i+1} P_z^{(i+1)} \Big|_{S_{i,i+1}}, \quad a_{i+1} P_z^{(i+1)} - g_z^{(i+1)} \frac{\partial P_z^{(i+1)}}{\partial z} = b_{i,i+1} P_z^{(i)} \Big|_{S_{i,i+1}}. \qquad (2b)$$

Opposite signs before the polarization derivatives, $\frac{\partial P_z^{(i)}}{\partial z}$ and $\frac{\partial P_z^{(i+1)}}{\partial z}$, in Eq.(2b) are due to the opposite signs of the outer normal to the interfaces, i.e., $\vec{n}_{i,i+1} = -\vec{n}_{i+1,i}$ (see e.g., **Fig. 1**). The boundary conditions at the electrode surfaces are:

$$a_1 P_z^{(1)} + g_z^{(1)} \frac{\partial P_z^{(1)}}{\partial z} = 0 \Big|_{S_{1,e}}, \quad a_N P_z^{(N)} - g_z^{(N)} \frac{\partial P_z^{(N)}}{\partial z} = 0 \Big|_{S_{N,e}}. \qquad (2c)$$

The boundary conditions (2) for polarization are of the third type [17]. In the particular case of a bilayer ($N = 2$), we obtain the following four conditions:

$$a_1 P_z^{(1)} + g_z^{(1)} \frac{\partial P_z^{(1)}}{\partial z} - b_{1,2} P_z^{(2)} = 0 \Big|_{z=h_1}, \quad a_2 P_z^{(2)} - g_z^{(2)} \frac{\partial P_z^{(2)}}{\partial z} - b_{1,2} P_z^{(1)} = 0 \Big|_{z=h_1}, \qquad (3a)$$

$$a_1 P_z^{(1)} + g_z^{(1)} \left.\frac{\partial P_z^{(1)}}{\partial z}\right|_{z=0} = 0, \qquad a_2 P_z^{(2)} - g_z^{(2)} \left.\frac{\partial P_z^{(2)}}{\partial z}\right|_{z=h_1+h_2} = 0. \qquad (3b)$$

The sketch of the considered bilayer capacitor is shown in **Fig. 1**.

If random fields created by charged or elastic defects exist in the multilayer, the sign-alternating randomly distributed electric charge sources $\delta\rho(x, y, z)$ should be included in the right-hand side of the Poisson equations for the electric potential $\varphi^{(i)}$ inside each layer:

$$\varepsilon_0 \varepsilon_b^{(i)} \left(\frac{\partial^2}{\partial x^2} + \frac{\partial^2}{\partial y^2} + \frac{\partial^2}{\partial z^2}\right) \varphi^{(i)} = \frac{\partial^2 P_z^{(i)}}{\partial z^2} + \delta\rho(x, y, z). \qquad (4a)$$

The electric boundary conditions are the continuity of $\varphi^{(i)}$ and electric displacement component $D_z^{(i)}$ at the interfaces, and the fixed potential at the electrodes,

$$\varphi^{(i)} = 0\big|_{S_{1,e}}, \qquad \varphi^{(i)} = U\big|_{S_{N,e}}. \qquad (4b)$$

We put $\delta\rho = 0$ for the derivation of the single-domain analytical solutions and consider nonzero $\delta\rho$ as the source of domain nucleation in the finite-element method (FEM) calculations.

The parameters $a_i$ and $b_{i,i+1}$ are determined by the short-range interactions at the interface, which can in principle be estimated employing Density Functional Theory (DFT) calculations [18, 19]; however the estimate may significantly deviate from the parameters fitted to experiment because the strength and sign of the short-range surface interactions are strongly dependent on the experimental preparation method and surface conditions.

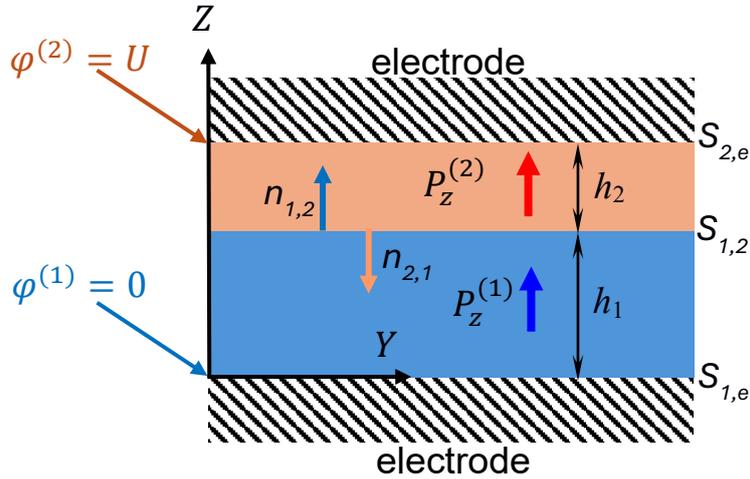

**FIGURE 1**. The sketch of the considered bilayer capacitor, where the layers have different thicknesses ($h_1$ and $h_2$) and polarizations $P_z^{(1)}$ (blue arrow) and $P_z^{(2)}$ (red arrow). The interface area between the layers is denoted as $S_{1,2}$, and the surface areas between the electrodes and the layers are denoted as $S_{1,e}$ and $S_{2,e}$, respectively. The "outer" normal from the interface 1 to the interface 2, denoted as $\vec{n}_{1,2}$, has is directed oppositely to the "outer" normal from the interface 2 to the interface 1, denoted as $\vec{n}_{2,1}$. The sample coordinates (X, Y, Z) form a right-handed coordinate system.

Hereinafter, we regard the electrodes and contacts as electrically "perfect", and such that imperfect screening conditions or dielectric gaps are absent at the electrode-layer interface. However, the long-range electric fields inevitably appear due to the depolarization effects related to the polarization difference, as well as domain formation in the layers. The field provides the long-range electrostatic interaction between the layers.

With respect to the short-range polarization coupling at the interface [18], the interfaces between the layers can be "matched", "mismatched" or "decoupled" depending on the short-range interaction constants $a_i$ and $b_{i,i+1}$. In the case of "decoupled" interfaces, the boundary conditions for polarization are natural at all interfaces ($\frac{\partial P_z^{(i)}}{\partial z} = 0$), precluding domain formation. In the case of a "matched" interface, the polarization is continuous at the layer interface ($P_z^{(i)} = P_z^{(i+1)}$) and the 2D and 3D domains may appear. The case of a "mismatched" interface corresponds to $P_z^{(i)} \neq P_z^{(i+1)}$ and/or $\frac{\partial P_z^{(i)}}{\partial z} \neq 0$ due to the nonzero coupling constants $b_{1,2}$ in Eq.(2b). Using FEM, we considered all three types of boundary conditions and conclude that they do not influence significantly the polar state of the layers thicker than 20 – 50 nm.

The case of decoupled interfaces allows analytical solutions that are presented next. For the case of domains being absent, when $\frac{\partial P_z^{(i)}}{\partial z} = 0$, the electric field inside the layers is given by expression

$$E_z^{(i)} = -\frac{P_z^{(i)} - \bar{D}}{\varepsilon_0 \varepsilon_b^{(i)}} + \frac{U}{\varepsilon_b^{(i)} \sum_{k=1}^{N}\left(h_k/\varepsilon_b^{(k)}\right)}. \qquad (5a)$$

Here $\bar{D}$ is the average displacement of the multilayer, which is equal to [20]:

$$\bar{D} = \left(\sum_{i=1}^{N} \frac{h_i}{\varepsilon_b^{(i)}} \bar{P}_z^{(i)}\right) \bigg/ \left(\sum_{k=1}^{N} \frac{h_k}{\varepsilon_b^{(k)}}\right), \qquad (5b)$$

where $\bar{P}_z^{(i)}$ is the average polarization of the layer "$i$". Here, $U$ is the electric potential applied across the multilayer.

The electric field in Equation (5a) has the external contribution, $\frac{U}{\varepsilon_b^{(i)} \sum_{k=1}^{N}\left(h_k/\varepsilon_b^{(k)}\right)}$, and the depolarization [21] contribution, $-\frac{P_z^{(i)} - \bar{D}}{\varepsilon_0 \varepsilon_b^{(i)}}$. Note that the average displacement $\bar{D}$ coincides with the average polarization $\bar{P}$ when $\varepsilon_b^{(i)}$ is the same for all layers. For simplicity, we consider uniform $\varepsilon_b^{(i)}$ in this work. The assumption of uniform $\varepsilon_b^{(i)}$ seems to be reasonable for the multilayers consisting of similar materials, like $Al_{x-1}Sc_xN/AlN$, $Al_{1-x}B_xN/AlN$ and $Zn_{1-x}Mg_xO/ZnO$. The general case of nonuniform $\varepsilon_b^{(i)}$ will be studied elsewhere in future publications.

We neglected the electrostriction coupling in this work, assuming that elastic stresses and strains have no significant impact in single-domain multilayers deposited on elastically matched substrates. However, we would like to underline that the elastic subsystem contribution should be very important for the polarization switching kinetics, because the inhomogeneous stresses and/or strains are accumulated at the domain walls [22, 23]. We defer the important issue for further studies.

### 3. Proximity-induced ferroelectric switching in ferroelectric bilayer capacitors

Let us consider a bilayer capacitor with different thickness of the layers (including, $h_1 \gg h_2$, $h_1 \sim h_2$ and $h_1 \ll h_2$). Hereinafter we regard that the first layer "1" is made of a "reference" ferroelectric material that is switchable and robust (e.g., $Al_{x-1}Sc_xN$ or $Zn_{1-x}Mg_xO$). It has a high spontaneous polarization $P_S$ (above 100 – 150 μC/cm$^2$) and thermodynamic coercive field $E_c$ (above 6 – 12 MV/cm), which are experimentally observable variables [2, 4, 5, 13, 24]. The double-well free energy of the reference ferroelectric "1" is shown in **Fig. 2(a)** by the black curve (the depth of the wells is -0.5 GPa or deeper). The second layer "2" can be a paraelectric material (such as quantum paraelectric $Eu_xSr_{1-x}TiO_3$, or incipient ferroelectrics like $KTaO_3$ and $TiO_2$ (rutile) [25, 26], or a dielectric material (such as or MgO, or $Al_2O_3$), whose parabolic potential well is shown by the orange curve in **Fig. 2(a)**. Also, the second layer "2" can be a "weak" (e.g., relaxor-like) ferroelectric with a small spontaneous polarization and coercive field, whose double-well potential is very shallow (e.g., shallower than -0.1 GPa, as shown by the brown curve in **Fig. 2(a)**). The special interest of this work is the case when the second layer is made of a polar material (such as AlN), whose spontaneous polarization is large and the thermodynamic coercive field is very close to, or higher than the dielectric breakdown field of the layer in its bulk form (typically ~ 6 – 7 MV/cm). The double-well free energies of such materials are shown by the light-green and dark-cyan curves in **Fig. 2(a).** Since we do not consider the temperature dependences here, the only requirement for the material "2" is the parabolic well or a double-well potential at a fixed working temperature.

In **Appendix A** [27] we derived approximate analytical expressions showing how the electric field (5) renormalizes the coefficients $\alpha_i$, $\beta_i$ and $\gamma_i$ in the TDLGD Eqs.(2a). Namely, the coefficients transform ("→") into the expressions:

$$\alpha_1 \to \alpha_1^{(R)} \approx \frac{\alpha_2 \aleph_1 + \alpha_1 \alpha_2 + \alpha_1 \aleph_2}{\aleph_1}, \qquad \alpha_2 \to \alpha_2^{(R)} \approx \frac{\alpha_2 \aleph_1 + \alpha_1 \alpha_2 + \alpha_1 \aleph_2}{\aleph_2}, \tag{6a}$$

$$\beta_1 \to \beta_1^{(R)} \approx \beta_2 \left\{1 + \frac{\alpha_1}{\aleph_1}\right\}^3 + \frac{\beta_1(\alpha_2 + \aleph_2)}{\aleph_1}, \qquad \beta_2 \to \beta_2^{(R)} \approx \beta_1 \left\{1 + \frac{\alpha_2}{\aleph_2}\right\}^3 + \frac{\beta_2(\alpha_1 + \aleph_1)}{\aleph_2}, \tag{6b}$$

$$\gamma_1 \to \gamma_1^{(R)} \approx \gamma_2 \left\{1 + \frac{\alpha_1}{\aleph_1}\right\}^5 + \frac{\gamma_1(\alpha_2 + \aleph_2)}{\aleph_1} + \left\{1 + \frac{\alpha_1}{\aleph_1}\right\}^2 \frac{3\beta_1 \beta_2}{\aleph_1}, \tag{6c}$$

$$\gamma_2 \to \gamma_2^{(R)} \approx \gamma_1 \left\{1 + \frac{\alpha_2}{\aleph_2}\right\}^5 + \frac{\gamma_2(\alpha_1 + \aleph_1)}{\aleph_2} + \left\{1 + \frac{\alpha_2}{\aleph_2}\right\}^2 \frac{3\beta_1 \beta_2}{\aleph_2}. \tag{6d}$$

Here the positive factors $\aleph_1$ and $\aleph_2$, originated from the depolarization field, are introduced:

$$\aleph_1 = \frac{h_2/\varepsilon_b^{(2)}}{\varepsilon_0 \varepsilon_b^{(1)}\left[h_1/\varepsilon_b^{(1)} + h_2/\varepsilon_b^{(2)}\right]}, \qquad \aleph_2 = \frac{h_1/\varepsilon_b^{(1)}}{\varepsilon_0 \varepsilon_b^{(2)}\left[h_1/\varepsilon_b^{(1)} + h_2/\varepsilon_b^{(2)}\right]}. \tag{7}$$

The accuracy of expressions (6) is high when $\aleph_1$ and $\aleph_2$ are higher than $10^{-2}$, and they become invalid if $\aleph_1$ or $\aleph_2$ tends to zero. Algebraic expressions (6) are relatively simple, they allow several important conclusions.

Let us consider the case when the layer "2" is a linear dielectric, which corresponds to $\alpha_2 > 0$ and $\beta_2 = \gamma_2 = 0$, and the layer "1" is a second-order ferroelectric with $\alpha_1 < 0$, $\beta_1 > 0$ and $\gamma_2 \geq 0$. In this case, the "renormalized" coefficients $\alpha_2^{(R)}$ in Eq.(6a) may become negative, and the renormalized coefficient $\beta_2^{(R)} \approx \beta_1\left\{1 + \frac{\alpha_2}{\aleph_2}\right\}^3$ in Eq.(6b) appears positive, and $\gamma_2^{(R)} \approx \gamma_1\left\{1 + \frac{\alpha_2}{\aleph_2}\right\}^5$ in Eq.(6d) is positive (for $\gamma_1 > 0$) or zero (for $\gamma_1 = 0$). Hence the proximity of the ferroelectric layer may induce apparent ferroelectricity in the linear dielectric layer. The same situation is possible for the nonlinear dielectric/ferroelectric and paraelectric/ferroelectric stacks.

Another important case is $\aleph_1 \approx \aleph_2 \approx \aleph$, which is possible for $h_1 \approx h_2$. Since, as a rule $\frac{|\alpha_i|}{\aleph} \ll 1$ (due to the giant factor $\frac{1}{\varepsilon_0} \sim 10^{11}$ m/F in $\aleph$), the expressions (6) acquire a simple form, namely $\alpha_1^{(R)} \approx \alpha_2 + \alpha_1 \approx \alpha_2^{(R)}$, $\beta_1^{(R)} \approx \beta_2 + \beta_1 \approx \beta_2^{(R)}$ and $\gamma_1^{(R)} \approx \gamma_1 + \gamma_2 + \frac{3\beta_1\beta_2}{\aleph} \approx \gamma_2^{(R)}$ with a high accuracy. This means that nominally different layers of close thickness become almost indistinguishable in the bilayer stack, and their polarization switching is almost the same (including the same coercive fields, remanent polarizations, and the shape of loops). Below we will demonstrate that this result remains valid for a much wider range of the layers thickness ratio $\frac{h_2}{h_1}$.

Actually, for the common case $\frac{|\alpha_i|}{\aleph_i} \ll 1$ and arbitrary ratio $\frac{\aleph_1}{\aleph_2} = \frac{h_2}{h_1}$, we obtained $\alpha_1^{(R)} \approx \frac{\alpha_2\aleph_1 + \alpha_1\alpha_2 + \alpha_1\aleph_2}{\aleph_1}$, $\beta_1^{(R)} \approx \frac{\beta_2\aleph_1 + \beta_1\aleph_2}{\aleph_1}$, $\gamma_1^{(R)} \approx \frac{\gamma_2\aleph_1 + \gamma_1\aleph_2 + 3\beta_1\beta_2}{\aleph_1}$, $\alpha_2^{(R)} \approx \frac{\alpha_2\aleph_1 + \alpha_1\alpha_2 + \alpha_1\aleph_2}{\aleph_2}$, $\beta_2^{(R)} \approx \frac{\beta_1\aleph_2 + \beta_2\aleph_1}{\aleph_2}$, and $\gamma_2^{(R)} \approx \frac{\gamma_1\aleph_2 + \gamma_2\aleph_1 + 3\beta_1\beta_2}{\aleph_2}$. Thus the ratios of the corresponding renormalized coefficients are the almost same, namely $\frac{\alpha_1^{(R)}}{\alpha_2^{(R)}} \approx \frac{\beta_1^{(R)}}{\beta_2^{(R)}} \approx \frac{\gamma_1^{(R)}}{\gamma_2^{(R)}} \approx \frac{\aleph_2}{\aleph_1} \approx \frac{h_1}{h_2}$. This means that the remanent polarizations of the layers $P_S^{(i)}$ are almost the same with an accuracy of the ratio $\frac{|\alpha_i|}{\aleph_i}$; and the thermodynamic coercive fields $E_c^{(i)}$ relate as $\frac{E_c^{(1)}}{E_c^{(2)}} \approx \frac{\aleph_2}{\aleph_1} \approx \frac{h_1}{h_2}$ (see Eqs.(A.13b) in **Appendix A** [27]). The equations for the polarization of the different layers, also derived in **Appendix A** [27], are:

$$\alpha_1^R P_z^{(1)} + \beta_1^{(R)}\left(P_z^{(1)}\right)^3 + \gamma_1^{(R)}\left(P_z^{(1)}\right)^5 \approx \eta_1^{(R)} E_a, \tag{8a}$$

$$\alpha_2^R P_z^{(2)} + \beta_2^{(R)} \left(P_z^{(2)}\right)^3 + \gamma_2^{(R)} \left(P_z^{(2)}\right)^5 \approx \eta_2^{(R)} E_a, \quad (8b)$$

where $\eta_1^{(R)} \approx \eta_2 + \frac{\aleph_2}{\aleph_1}\eta_1$ and $\eta_2^{(R)} \approx \eta_1 + \frac{\aleph_1}{\aleph_2}\eta_2$ are the "splitting" factors of the applied field, and $E_a = \frac{U}{h_1+h_2}$ is the field applied to the bilayer capacitor. More rigorous cumbersome expressions for the splitting factors $\eta_i^{(R)}$ are given by Eq.(A.9e-f) in **Appendix A** [27]. Due to the field splitting, the "effective" coercive fields of the layers are $E_{ca}^{(i)} = \frac{E_c^{(i)}}{\eta_i^{(R)}}$. Their ratio is unity, namely $\frac{E_{ca}^{(1)}}{E_{ca}^{(2)}} = \frac{\eta_2^{(R)}}{\eta_1^{(R)}} \frac{E_c^{(1)}}{E_c^{(2)}} \approx 1$, because $\frac{\eta_2^{(R)}}{\eta_1^{(R)}} \approx \frac{\aleph_1}{\aleph_2}$ and $\frac{E_c^{(1)}}{E_c^{(2)}} \approx \frac{\aleph_2}{\aleph_1}$. Thus, the effective coercive fields in both layers are actually the same, $E_{ca}^{(1)} \approx E_{ca}^{(2)}$ (see Eqs.(A.13c) in **Appendix A** [27]). Hence one can expect the "simultaneous" collective switching in the ferroelectric/ferroelectric, ferroelectric/paraelectric and ferroelectric/dielectric bilayers, with the relative dielectric constant of the dielectric being 10 or higher.

To consider a general case, let us introduce the thickness ratio and the dimensionless variables in the free energy given in Eq.(1). Dimensionless variables and parameters are summarized in **Table I**. Here the dimensionless polarization $\tilde{P}$ in both layers is normalized to the characteristic polarization of the first layer, $P_c \cong \sqrt{-\alpha_1/\beta_1}$, and the electric field $\tilde{E}$ is normalized to the characteristic field, $E_c \cong -\alpha_1 P_c$, of the same layer. The dimensionless LGD expansion coefficients, $\tilde{\alpha}$, $\tilde{\beta}$, and $\tilde{\gamma}_i$, are normalized to the parameters of the first layer. The introduction of dimensionless values is valid for the case of $\beta_1 > 0$ and $\alpha_1 < 0$. If $\gamma_1 = 0$ the values $P_c$ and $E_c$ are respectively the spontaneous polarization and the thermodynamic coercive field (with the factor $2/3\sqrt{3}$) of the first layer. Two other dimensionless parameters, named the depolarization factors $\xi_i$, originating from the depolarization field in Eq.(5a), are rather small on the order of $10^{-2}$. The higher the $\xi_i$, the larger the contribution of the depolarization field that couples the layers.

**Table I.** Dimensionless variables and parameters (valid only for $\beta_1 > 0$ and $\alpha_1 < 0$).

| Thickness ratio | Polarization | Electric field | LGD expansion coefficients | | | Depolarization factors |
|---|---|---|---|---|---|---|
| $\tilde{h} = \frac{h_2}{h_1}$ | $\tilde{P} = P_z\sqrt{-\frac{\beta_1}{\alpha_1}}$ | $\tilde{E} = \frac{E_z}{-\alpha_1}\sqrt{-\frac{\beta_1}{\alpha_1}}$ | $\tilde{\alpha} = -\frac{\alpha_2}{\alpha_1}$ | $\tilde{\beta} = \frac{\beta_2}{\beta_1}$ | $\tilde{\gamma}_1 = -\alpha_1\frac{\gamma_1}{\beta_1^2}$ | $\xi_1 = \varepsilon_0\varepsilon_b^{(1)}|\alpha_1|$ |
| | | | | | $\tilde{\gamma}_2 = -\alpha_1\frac{\gamma_2}{\beta_1^2}$ | $\xi_2 = \varepsilon_0\varepsilon_b^{(2)}|\alpha_1|$ |

For the sake of simplicity in this section we consider the model case $\tilde{\gamma}_1 = \tilde{\gamma}_2 = 0$ and $\xi_1 = \xi_2$, which corresponds to the bilayer made of two similar materials (e.g., pure and doped compounds) with the true (or apparent) second order phase transition at $\alpha_1 = 0$ and $\alpha_2 = 0$. Unfortunately, the analysis becomes very complex for nonzero $\tilde{\gamma}_i$, because this adds two extra parameters to the multi-parameter space. In **Appendix A** [27] we derive the conditions for the stable phase states of the bilayer, which are valid for the case of nonzero $\tilde{\gamma}_i$.

The dimensionless spontaneous polarization of the reference bulk ferroelectric "1" is equal to unity, the thermodynamic coercive field is equal to $2/3\sqrt{3}$ (about 0.4), and the free energy minimum is equal to $-\frac{1}{4}$ at $E = 0$. The black curve "FE" in **Fig. 2(a)** corresponds to the bulk ferroelectric material with $\tilde{\alpha} = -1$ and $\tilde{\beta} = 1$, which acts as the layer "1" of finite thickness $h_1$ in the calculations presented below. The layer "2" has several choices for the potential well with different values of the dimensionless coefficients $\tilde{\alpha}$ and $\tilde{\beta}$, and different layers thickness ratio $\tilde{h}$ (from 0.1 to 10) are shown in **Fig. 2(a)**. Various combinations of such bilayers are discussed next.

*Ferroelectric (FE, layer 1) / Paraelectric (PE, layer 2) bilayer stack*: The physical picture shown in **Fig. 2(b)** corresponds to the case when the first layer of thickness $h_1$ is made of the reference ferroelectric "FE", and the second layer of thickness $h_2$ is made of a paraelectric material (with $\tilde{\alpha} = +1$ and $\tilde{\beta} = 1$), whose parabolic potential is shown by the orange curve "PE" in **Fig. 2(a)**. Polarization hysteresis loop is absent for $\tilde{h} > 1$ (see magenta and red curves in **Fig. 2(b)**), because the thicker paraelectric layer destroys the ferroelectric order in the thinner ferroelectric layer due to the strong depolarization field (given by Eqs.(5)). As a result, the renormalized coefficient $\alpha_1^{(R)}$ becomes positive, while $\alpha_1$ is negative (see Eq.(6a)). Polarization hysteresis loop appears for $\tilde{h} \approx 1$ and becomes significantly higher and wider with a decrease in $\tilde{h}$ (see black, green and blue curves in **Fig. 2(b)**), because the thinner paraelectric layer cannot destroy the ferroelectric order in the thicker ferroelectric layer. Due to the proximity effect (i.e., the field coupling) the polarization of the paraelectric layer also undergoes the ferroelectric switching for $\tilde{h} < 1$ (see solid and dashed curves in **Fig. A1(a)** placed in the end of **Appendix A**). Thus, the paraelectric layer becomes a ferroelectric, a striking case of proximity ferroelectricity that has not yet been demonstrated experimentally but is being predicted by this theory. The same prediction can be made for the dielectric/ferroelectric stack.

*Ferroelectric (FE, layer 1) / Weak-ferroelectric (W-FE, layer 2) bilayer stack*: The situation shown in **Fig. 2(c)** corresponds to the case when the first layer of thickness $h_1$ is made of the ferroelectric "FE", and the second layer of thickness $h_2$ is made of a "weak" ferroelectric ($\tilde{\alpha} = -0.1$, $\tilde{\beta} = 1$), with a shallow double-well potential as shown by the brown curve "W-FE" in **Fig. 2(a)**. Polarization hysteresis loop has a small remanent polarization ($\tilde{P}_S \ll 1$) and a low coercive field ($\tilde{E}_c \ll 1$) for $\tilde{h} \gg 1$ (see magenta and red curves in **Fig. 2(c)**), because the thick "weak" ferroelectric layer dominates over the thin normal ferroelectric layer due to the depolarization field induced in the layers. Polarization hysteresis loop becomes significantly higher and wider with decrease in $\tilde{h}$ (see black, green and blue curves in **Fig. 2(c)**), because the strong ferroelectric order in the thicker layer begins to dominate in the average polarization and decreases the depolarization field in a self-consistent way. In result the polarization in both layers undergoes the collective ferroelectric switching, namely the remanent values of $P_z^{(1)}$ and $P_z^{(2)}$, as well as corresponding coercive fields and loop shape, almost coincide (compare solid and dashed curves in **Fig. A1(b)** in **Appendix A**).

*Ferroelectric (FE, layer 1) / Hard-ferroelectric (H-FE, layer 2) bilayer stack*: The physical picture shown in **Fig. 2(d)** illustrates the case when the first layer of thickness $h_1$ is made of the ferroelectric "FE" and the second layer of thickness $h_2$ is made of the hard ferroelectric material ($\tilde{\alpha} = -1, \tilde{\beta} = 0.1$) with a large coercive field, whose deep and wide double-well potential is shown by the dark-cyan curve "H-FE" in **Fig. 2(a).** The remanent polarization is very high ($\tilde{P}_s > 2$) and the thermodynamic coercive field ($\tilde{E}_c \geq 0.8$) may appear close to the electric breakdown field for $\tilde{h} \gg 1$ (see magenta and red curves in **Fig. 2(d)**), because the polar properties of the thick hard ferroelectric layer dominates over the properties of the thin normally-switchable ferroelectric layer. However, the depolarization field tends to reduce the effective coercive field (resulting in $\tilde{E}_{c1} < \tilde{E}_c < \tilde{E}_{c2}$) and to equalize the polarizations of layers (resulting in $P_z^{(1)} \approx P_z^{(2)}$). As a result, we observe the collective switching picture for the case $\tilde{h} = 1$ (see black curve in **Fig. 2(d)**). Polarization hysteresis loop becomes significantly lower and narrower with further decrease in $\tilde{h}$ (see green and blue curves in **Fig. 2(d)**), because the polar properties of the thicker normally-switchable ferroelectric layer begin to dominate over the properties of the thinner H-FE layer. It is important to note that both layers undergoes the collective ferroelectric switching in a similar way, namely the hysteresis loops of polarizations $P_z^{(1)}$ and $P_z^{(2)}$ have a very similar shape, the same coercive fields and close values of the remanent polarizations (compare solid and dashed curves in **Fig. A1(c)** in **Appendix A**).

*Ferroelectric (FE, layer 1) / Non-ferroelectric (N-FE, layer 2) bilayer stack*: The loops shown in **Fig. 2(e)** correspond to the case when the first layer of thickness $h_1$ is made of the ferroelectric "FE" and the second layer of thickness $h_2$ is made of the non-switchable polar material ($\tilde{\alpha} = -10, \tilde{\beta} = 10$), whose deep and narrow double-well potential is shown by the light-green curve "N-FE" in **Fig. 2(a).** The dielectric permittivity of the non-switchable polar material $\varepsilon_2 \sim \frac{1}{|\alpha_2|}$ is much smaller than the ferroelectric permittivity $\varepsilon_1 \sim \frac{1}{|\alpha_1|}$ of the thin normally-switchable ferroelectric, because $-\frac{1}{\alpha_1} \gg -\frac{1}{\alpha_2}$. Due to the splitting of external field inside the layers, which can be estimated as $\frac{E_z^{(2)}}{E_z^{(1)}} \sim \frac{\varepsilon_1}{\varepsilon_2} \gg 1$, we observe the significant reduction in the *effective* coercive field $\tilde{E}_c$ with decrease in $\tilde{h}$ (compare red, magenta and black curve with $\tilde{E}_c \approx 2$ in **Fig. 2(e)**). Polarization hysteresis loop becomes significantly narrower with further decrease in $\tilde{h}$ (see green and blue curves in **Fig. 2(e)**), because the dielectric properties of the thicker normally-switchable ferroelectric layer begin to dominate over the properties of the thinner non-switchable polar layer. Similarly to the previous case (shown in **Fig. A1(c)**) both layers undergo the collective ferroelectric switching (compare solid and dashed curves in **Fig. A1(d)** in **Appendix A**). This means that the reduction in the effective coercive field of the second N-FE layer is due to the changes in its double-well potential. However, the physical origin of the potential change is the coupling of the layers by the internal electric fields.

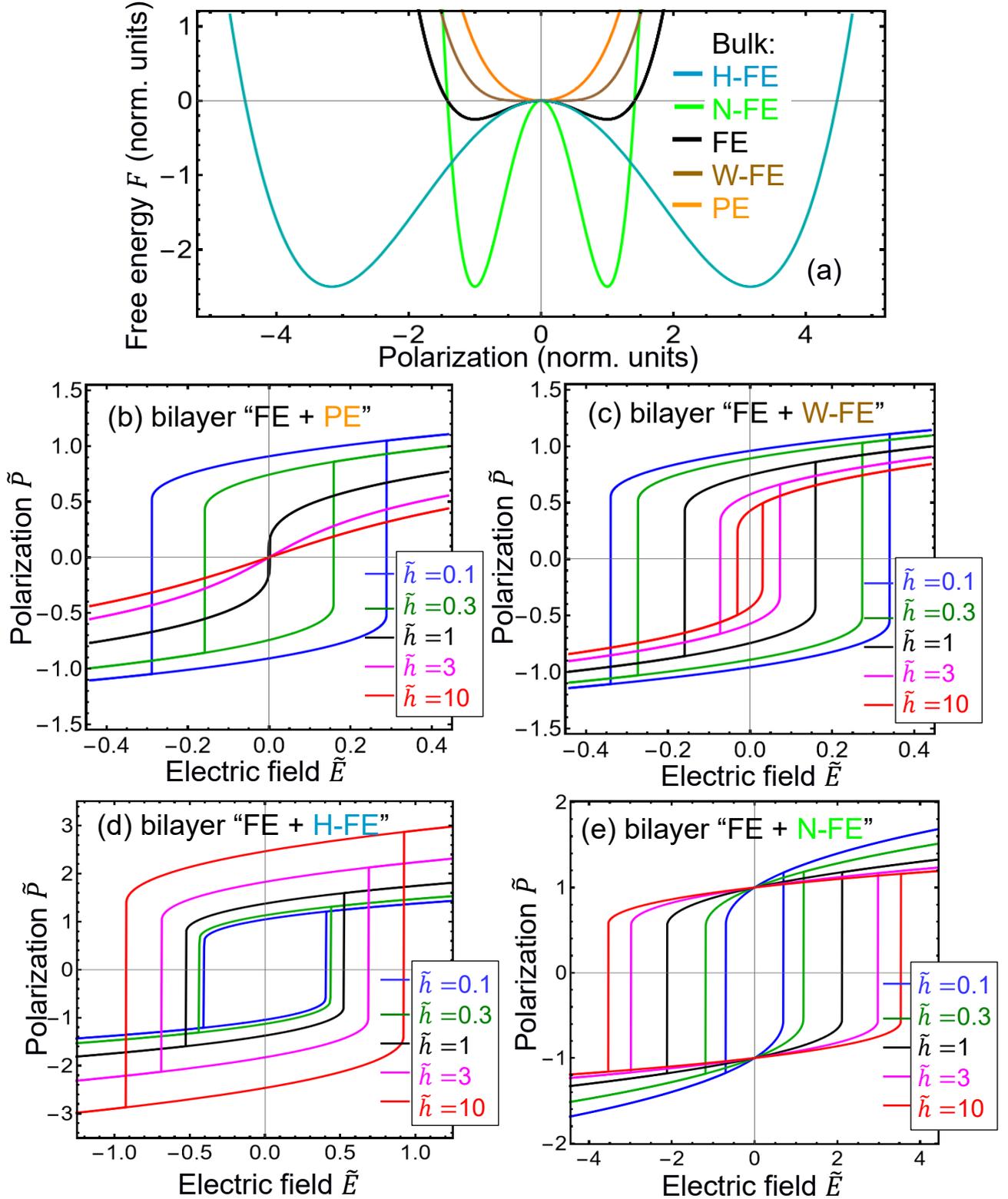

**FIGURE 2. (a)** The sketch of the free energy of the bulk materials. The black curve corresponds to the reference

ferroelectric (FE) with $\tilde{\alpha} = -1$ and $\tilde{\beta} = 1$, the orange curve corresponds to the paraelectric (PE) with $\tilde{\alpha} = +1$ and $\tilde{\beta} = 1$, the brown curve corresponds to the weak ferroelectric (W-FE) with $\tilde{\alpha} = -0.1$ and $\tilde{\beta} = 1$. Other curves correspond to the hardly switchable ferroelectrics (H-FE) with $\tilde{\alpha} = -1$, $\tilde{\beta} = 0.1$ (the dark-cyan curve) and non-switchable ferroelectrics (N-FE) $\tilde{\alpha} = -10$, $\tilde{\beta} = 10$ (the light-green curve). The energy scale corresponds to GPa, and the polarization scale corresponds to ~100 µC/cm². **(b)-(e)** Hysteresis loops of bilayer polarization calculated for the fixed coefficients of the first layer ($\tilde{\alpha} = -1$, $\tilde{\beta} = 1$) and different coefficients of the second layer $\tilde{\alpha} = +1$, $\tilde{\beta} = 1$ **(b)**, $\tilde{\alpha} = -0.1$, $\tilde{\beta} = 1$ **(c)**, $\tilde{\alpha} = -1$, $\tilde{\beta} = 0.1$ **(d)**, and $\tilde{\alpha} = -10$, $\tilde{\beta} = 10$ **(e)**. The thickness ratio $\tilde{h}$=0.1, 0.3, 1, 3 and 10 (blue, green, black, magenta and red curves, respectively). Other parameters are $\tilde{\gamma}_1 = \tilde{\gamma}_2 = 0$ and $\xi_1 = \xi_2 = 0.02$.

Note that the apparent reversibility of the bilayer consisting of the switchable ferroelectric and non-switchable polar layer of close thickness (see **Fig. 2(e)**) can be classified as a pronounced proximity-induced ferroelectricity. Below we analyze the range of LGD parameters $\tilde{\alpha}$, $\tilde{\beta}$ and thickness ratio $\tilde{h}$ for which the proximity-induced ferroelectricity may be observed.

The dependences of the bilayer remanent polarization $\tilde{P}_s$ and intrinsic coercive field $\tilde{E}_c$ on the dimensionless LGD coefficients $\tilde{\alpha}$ and $\tilde{\beta}$, and the thickness ratio $\tilde{h}$ are shown in **Fig. 3**. Dark-red regions correspond to the non-ferroelectric state (N-FE) with the very high irreversible spontaneous polarization ($\tilde{P}_s > 4$) and ultra-high "hypothetic" coercive field ($\tilde{E}_c > 4$) which exceeds significantly the electric breakdown field. Blue regions correspond to the paraelectric (PE) phase. The gradual change of the color from orange to light yellow corresponds to the appearance of the proximity-induced ferroelectric (Pr-FE) state. The gradual change of the color from light yellow to light blue corresponds to the gradual transition from the Pr-FE state to the normal ferroelectric (FE) phase.

The N-FE states exist for the high negative values of $\tilde{\alpha}$ and small or moderate values of $\tilde{\beta}$, which are required for the high value of $\tilde{E}_{c2} \sim -\tilde{\alpha}\sqrt{-\tilde{\alpha}/\tilde{\beta}}$ (see **Fig. 3(a)** and **3(d)** calculated for $\tilde{h} = 1$ and **Fig. B1** [27] calculated for $10 \leq \tilde{h} \leq 0.1$). The large region of the Pr-FE states separates the smaller region of N-FE states from the normal FE phase for the ratio $\tilde{h} = 1$ (see **Fig. 3(a)** and **3(d)**). The almost vertical boundary between the FE and the PE phase implies the independence of the FE-PE transition on the $\tilde{\beta}$ value. The $\tilde{\alpha}$-position of the FE-PE boundary very strongly shifts to the left with increase in $\tilde{h}$, because positive and big $\tilde{\alpha}$ favors the PE state in the second layer (see **Fig. B1** [27]).

In the dependence on the ratio $\tilde{h}$ and coefficient $\tilde{\alpha}$ the bilayer can be in the N-FE state (for $\tilde{h} \gg 1$ and $\tilde{\alpha} \ll -1$), or in the PE state (for $\tilde{h} \leq 1$ and $\tilde{\alpha} > 0$). The large region of the Pr-FE states separates the N-FE states from the PE phase (see **Fig. 3(b)** and **3(e)** calculated for $\tilde{\beta} = 0.1$ and **Fig. B2** [27] calculated for $10 \leq \tilde{\beta} \leq 0.1$). The boundary between the N-FE and Pr-FE states (as well as the boundary between the Pr-FE states and the PE phase) is curved due to the relatively strong influence of $\tilde{h}$ on the transition conditions, which is a typical manifestation of the size effect. Notably that bilayer switching appearing for $\tilde{h} > 1$, $\tilde{\alpha} \ll -1$ and $\tilde{\beta} \gg 1$ (see e.g., **Fig. 3(c)** and **3(f)** calculated for $\tilde{\alpha} = -10$), is a pronounced manifestation of the proximity-induced ferroelectric switching. The region of the Pr-FE state increases with increase in $\tilde{\beta}$ and decreases with an increase in $\tilde{h}$, because $\tilde{\beta} \gg 1$ weakens the polarization and coercive field of the non-switchable layer.

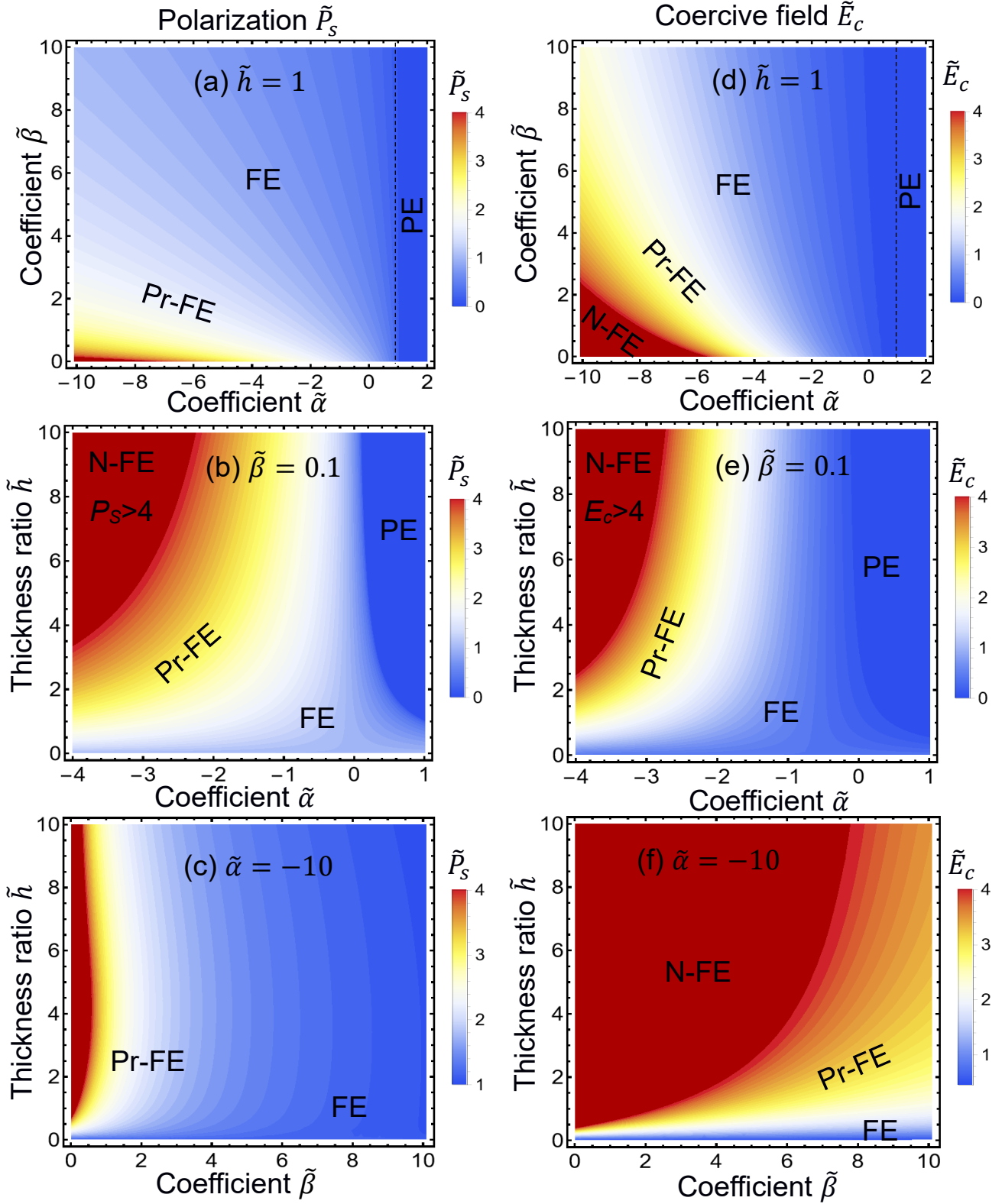

**FIGURE 3.** The dependences of the dimensionless remanent polarization $\tilde{P}_S$ (**a, b, c**) and intrinsic coercive field

$\tilde{E}_c$ **(d, e, f)** of the bilayer polarization hysteresis loops on the dimensionless LGD coefficients $\tilde{\alpha}$ and $\tilde{\beta}$, and thickness ratio $\tilde{h}$. The ratio $\tilde{h} = 1$ for the plots **(a)** and **(d)**, $\tilde{\beta} = 0.1$ for the plots **(b)** and **(e)**, and $\tilde{\alpha} = -10$ for the plots **(c)** and **(f)**. Other parameters are the same as in **Fig. 2**. Dark-red regions correspond to the non-ferroelectric state (N-FE) with the very high irreversible spontaneous polarization ($\tilde{P}_s > 4$) and ultra-high "hypothetic" coercive field ($\tilde{E}_c > 4$) which overcome the electric breakdown field. Blue regions correspond to the paraelectric (PE) phase. The color change from orange to light yellow corresponds to the proximity-induced ferroelectric (Pr-FE) state. The color change from the light yellow to the light blue corresponds to the gradual transition to the normal ferroelectric (FE) phase.

Note that hysteresis loops can also be observed for positive $\tilde{\alpha}$ (e.g., for $0 < \tilde{\alpha} \leq 1$), which may be interpreted as another manifestation of the proximity-induced ferroelectric switching effect in the paraelectric layer (see also **Fig. 2(b)**) and deserves a separate study for e.g., TiO$_2$/Ba$_{1-x}$Sr$_x$TiO$_3$ or MgO/Zn$_{1-x}$Mg$_x$O stacks. However, in this work we focus on the proximity-induced ferroelectric transition in the non-switchable layer with high thermodynamic coercive field well above the electric breakdown field.

### 4. Proximity-induced ferroelectricity in the AlN/Al$_{x-1}$Sc$_x$N bilayers

The sketch of the considered bilayer Al$_{x-1}$Sc$_x$N/AlN is shown in **Fig. 4(a)**. **Figure 4(b)** shows the free energy wells of the separate layers, where a shallow red well corresponds to the Al$_{0.73}$Sc$_{0.27}$N and a much deeper and wider blue well corresponds to AlN. LGD parameters of Al$_{0.73}$Sc$_{0.27}$N and AlN layers, which are listed in **Table II,** were determined from the experimentally measured spontaneous polarization [28, 29] and linear dielectric permittivity [30] as described in Ref. [31] and **Appendix C** [27]. The background permittivity $\varepsilon_b$ is estimated as the square of refractive index, which is about 2.1-2.3 for Al$_{0.73}$Sc$_{0.27}$N according to Ref. [32].

Table II. LGD model parameters for Al$_{x-1}$Sc$_x$N and AlN compounds

| compound | $\alpha$, m/F | $\beta$, m$^5$/(F C$^2$) | $\gamma$, m$^7$/(F C$^4$) | property |
|---|---|---|---|---|
| Al$_{0.73}$Sc$_{0.27}$N | $-2.644 \cdot 10^8$ | $-3.155 \cdot 10^9$ | $2.788 \cdot 10^9$ | switchable |
| AlN | $-2.164 \cdot 10^9$ | $-3.155 \cdot 10^9$ | $2.788 \cdot 10^9$ | non-switchable |

First, for the sake of simplicity, we again suppose the natural boundary conditions at all interfaces, $\frac{\partial P_i}{\partial z} = 0$, so that the polarization is homogeneous, $P_i \equiv \bar{P}_i$, and the driving force of the domain formation is absent. **Figure 4(c)** shows the thermodynamic single-domain polarization switching in the Al$_{0.73}$Sc$_{0.27}$N/AlN bilayer capacitor with the equal thickness of layers. Let us emphasize that the hysteresis loop corresponding to the single-domain switching of AlN capacitor cannot be observed experimentally, because the electric breakdown field $E_b$ of the pure AlN (which is about 6 - 8 MV/cm) is much lower than the calculated intrinsic coercive field, $E_c \approx 26$ MV/cm (see the blue loop in **Fig. 4(c)**). The hysteresis loop corresponding to the single-domain switching of AlScN capacitor has the giant intrinsic coercive field, $E_c \approx 9$ MV/cm, and so the single-domain switching of $Al_{0.73}Sc_{0.27}N$ is unlikely (see the red loop in **Fig. 4(c)**). Due to the same reasons, the hysteresis loop corresponding to the single-domain switching of the bilayer $Al_{0.73}Sc_{0.27}N$/AlN capacitor, whose coercive field is about 15 MV/cm (see the magenta loop in **Fig. 4(c)**), cannot be observed experimentally either. Despite thermodynamic coercive fields being giant, the effect of the AlN coercivity reduction is caused by the depolarization electric field, existing in both layers (see Eqs.(5)). The field tends to equalize the polarization values in both layers to minimize the depolarization field energy of the bilayer. Indeed, as one can see from **Fig. 4(d)**, the free energy minimum of the bilayer corresponds to $P_z^{(1)} \approx P_z^{(2)}$.

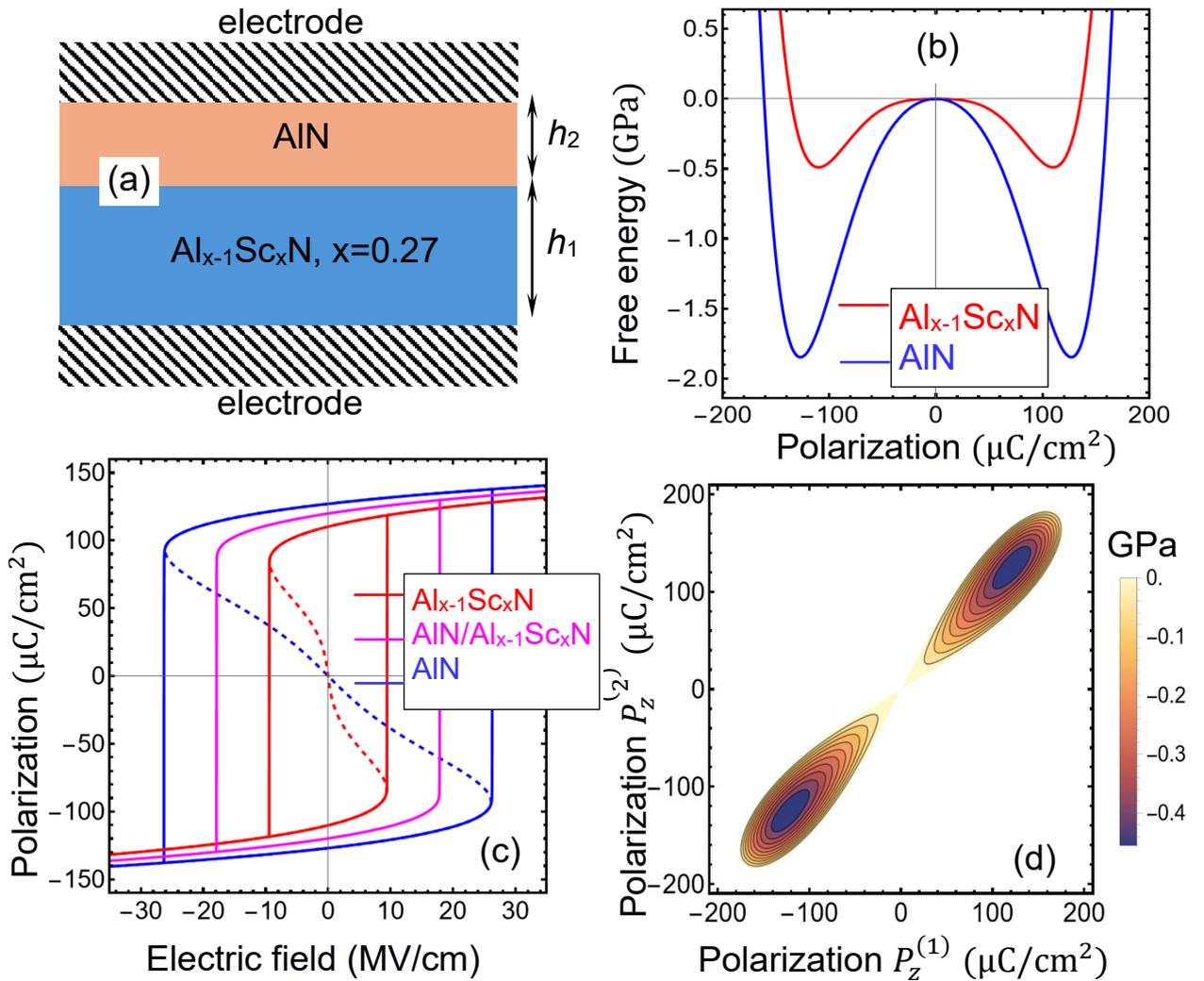

**FIGURE 4**. **(a)** The sketch of the considered bilayer $Al_{0.73}Sc_{0.27}N$/AlN. **(b)** Free energy wells of the separate layers, red and blue curves are for $Al_{0.73}Sc_{0.27}N$ and AlN, respectively. **(c)** The single-domain polarization hysteresis loops for the bulk $Al_{0.73}Sc_{0.27}N$ (the red loop), bulk AlN (the blue loop) and the AlN/$Al_{0.73}Sc_{0.27}N$

bilayer (the magenta loop). **(d)** The contour map of the bilayer capacitor free energy as a function of the layer polarizations $P_z^{(1)}$ and $P_z^{(2)}$. The thickness of each layer is 200 nm, $\varepsilon_b^{(1)} = \varepsilon_b^{(2)} = 4$. LGD parameters are listed in **Table II**.

The equalization of the layers' polarizations and effective coefficients $\alpha_{1,2}^{(R)}$ (which define the effective dielectric permittivity) may explain the experimentally observed [13] proximity-induced polarization switching in $Al_{x-1}Sc_xN/AlN$ bilayers. However, the reduction of the coercive field to the values close to those observed experimentally (e.g., below $E_b$) is required. In particular, the hysteresis loop, which corresponds to the polarization switching in the bilayer capacitor in the presence of charged defects, may be observed if the "resulting" coercive field is significantly smaller than the electric breakdown field due to the domain appearance at nucleation centers and domain intergrown through the bilayer.

To look for the possible sources of the domain appearance we considered various types of charged and elastic defects with different locations in the bilayer. Based on the results of FEM we conclude that randomly distributed electric charge sources $\delta\rho(x, y, z)$ in Eq.(4a) can significantly reduce the coercive field in a controllable manner if they are of a monopole type. The influence of random electric point dipoles on the coercive field appeared negligibly small. As anticipated the coercive field reduction depends strongly on the concentration of random charges in the bulk of thick layers ( ~ 100 nm or thicker), meanwhile the interface defects are significant for thin films ( ~ 10 nm or thinner).

An example of the electrostatic potential distribution in the $AlN/Al_{0.73}Sc_{0.27}N$ bilayer was calculated at $U = 0$ in the presence of randomly distributed pairs of positive and negative point charges (i.e., dipolar-type defects) is shown in **Fig. 5(a)**. Charged defects induce local electric fields, which aid in the switching acting as nucleation centers for emerging domains [33, 34]. The local fields induced by the dipolar-type defects are seen in **Fig. 5(a)**, and the defect-induced domain spots and needles are seen in **Fig. 6**. The average concentration of the defects $n_d$ is $7.8 \cdot 10^{25}$ m$^{-3}$ in the $Al_{0.83}Sc_{0.27}N$ layer and $8.2 \cdot 10^{25}$ m$^{-3}$ in the AlN layer. The average distance between positive and negative charge of the dipolar-type defect $l_d$ is about 1.5 nm, and the average charge $q_d$ is about $2.3e_0$ ($e_0$ is the elementary charge, $e_0 = 1.6 \cdot 10^{-19}$C). The average polarization of the dipolar-type defects is $P_d = n_d q_d l_d \cong 0.044$ C/m$^2$, that is much smaller than the spontaneous polarization of $Al_{0.83}Sc_{0.27}N$ and AlN layers (about $1 - 1.2$ C/m$^2$). We use slightly different concentrations of dipolar-type defects ($\delta n_d = 0.4 \cdot 10^{25}$ m$^{-3}$ or 5.1 %) in the $Al_{0.83}Sc_{0.27}N$ and AlN layers to reduce the coercive field of the layers to varying degrees: less in the $Al_{0.83}Sc_{0.27}N$ layer (coercive field reduction from 9 MV/cm to 4.9 MV/cm) than in the AlN layer (reduction from 26 MV/cm to 8.8 MV/cm) in trying to match experiment results [13] for Al$_{0.83}$Sc$_{0.27}$N/AlN bilayer. When we take equal $n_d = 8.0 \cdot 10^{25}$ m$^{-3}$ in both layers, it leads to the coercive fields about 12 MV/cm and 4 MV/cm in AlN and Al$_{0.83}$Sc$_{0.27}$N layers, respectively.

The reduction of the coercive field induced by the defects can be estimated from the widths of the polarization loops in **Fig. 5(b)**. Here the coercive field for each loop is significantly smaller than the corresponding thermodynamic coercive fields shown in **Fig. 4(c).** The loops in **Fig. 5(b)** are calculated by the FEM for the same concentration and location of charge defects in the capacitor, with the capacitor filled with the ferroelectric Al$_{0.73}$Sc$_{0.27}$N (the red loop), or by the non-switchable AlN (the blue loop), or by the Al$_{0.73}$Sc$_{0.27}$N and AlN layers of the same thickness (the magenta loop).

For calculations of bilayer polarization, we used the "matched" and "mismatched" boundary conditions at the interface (namely, we vary the dimensionless interfacial coupling constants $-1 \leq \tilde{b}_{1,2} \leq 1$ in Eq.(3a)). We conclude that the influence of $\tilde{b}_{1,2}$ is very short-range (being significant in ultra-thin bilayers with $h_1 + h_2 < 20$ nm) in comparison with the long-range influence of the randomly distributed electric charge sources $\delta\rho(x, y, z)$ inside the layers.

Comparison of the thermodynamic coercive fields in **Fig. 4(c)** with those reduced in the presence of defects and shown in **Fig. 5(b)**, demonstrates that the defects reduce the coercive field of AlN from about 26 MV/cm to 8.8 MV/cm (i.e., about 2.95 times). The coercive field of Al$_{0.83}$Sc$_{0.27}$N/AlN bilayer reduces from 18 MV/cm to 6.5 MV/cm (i.e., in 2.77 times); and the coercive field of Al$_{0.83}$Sc$_{0.27}$N layer reduces from 9 MV/cm to 4.9 MV/cm (i.e., in 1.88 times). Unequal reduction of the coercive field between AlN, Al$_{0.83}$Sc$_{0.27}$N/AlN and Al$_{0.83}$Sc$_{0.27}$N is observed. The ratios in the proportionality of the coercive fields without the defects is 26:18:9, which are higher than the corresponding ratios in the proportionality of 8.8:6.5:4.9 for switching with defects. We speculate that the switching is "more" extrinsic in AlN and "less" extrinsic in Al$_{0.83}$Sc$_{0.27}$N for specially chosen parameters of defects ($q_d$, $l_d$ and $n_d$). We do recognize that the parameters of defects in our study are convenient fitting parameters for modeling a charged defect. Also, the dependence of the observable coercive field on the defect concentration is nonlinear. The smaller the thermodynamic coercive field is, the less important defects are, because the system is closer to the intrinsic switching scenario. This observation can explain why the giant thermodynamic coercive field of AlN is reduced stronger (in 3 times) by approximately the same concentration of defects than those in Al$_{0.83}$Sc$_{0.27}$N, which thermodynamic coercive field is significantly smaller (and so the defects reduce it in 2 times only).

Therefore, nonlinear interaction of the intrinsic and extrinsic factors (via the nonlinear time-dependent LGD equations) leads to the situation when strongly different Landau expansion coefficients of AlN and Al$_{0.83}$Sc$_{0.27}$N together with slightly different concentrations of dipolar-type defects lead to about 3 times reduction in the coercive field of the AlN, and about 2 times reduction in the coercive field of the Al$_{0.83}$Sc$_{0.27}$N.

**Figures 5(c)** and **5(d)** illustrate the dependences of the bilayer remanent polarization $P_s$ and effective coercive field $E_c$ on the thickness $h_2$ of the AlN layer for the fixed thickness of $Al_{0.83}Sc_{0.27}N$ layer in the presence of dipolar-type charged defects. It is seen that $P_s$ and $E_c$ increase monotonically with increase in $h_2$. The bilayer structure is switchable for thickness $h_2 \leq h_{cr}$, where the value of the critical thickness $h_{cr}$ is determined by the breakdown coercive field $E_b$ (we choice it equal to 7 MV/cm), and by the concentration of random defects $\delta\rho(x,y,z)$. We treat the amplitude and concentration of the defects as fitting parameters, which are the close in both layers (even though it maybe not so), and vary them to reach the bilayer critical thickness $h_{cr} \geq h_1$ and to obtain the $P_s$ values close to the values experimentally measured in Ref. [13].

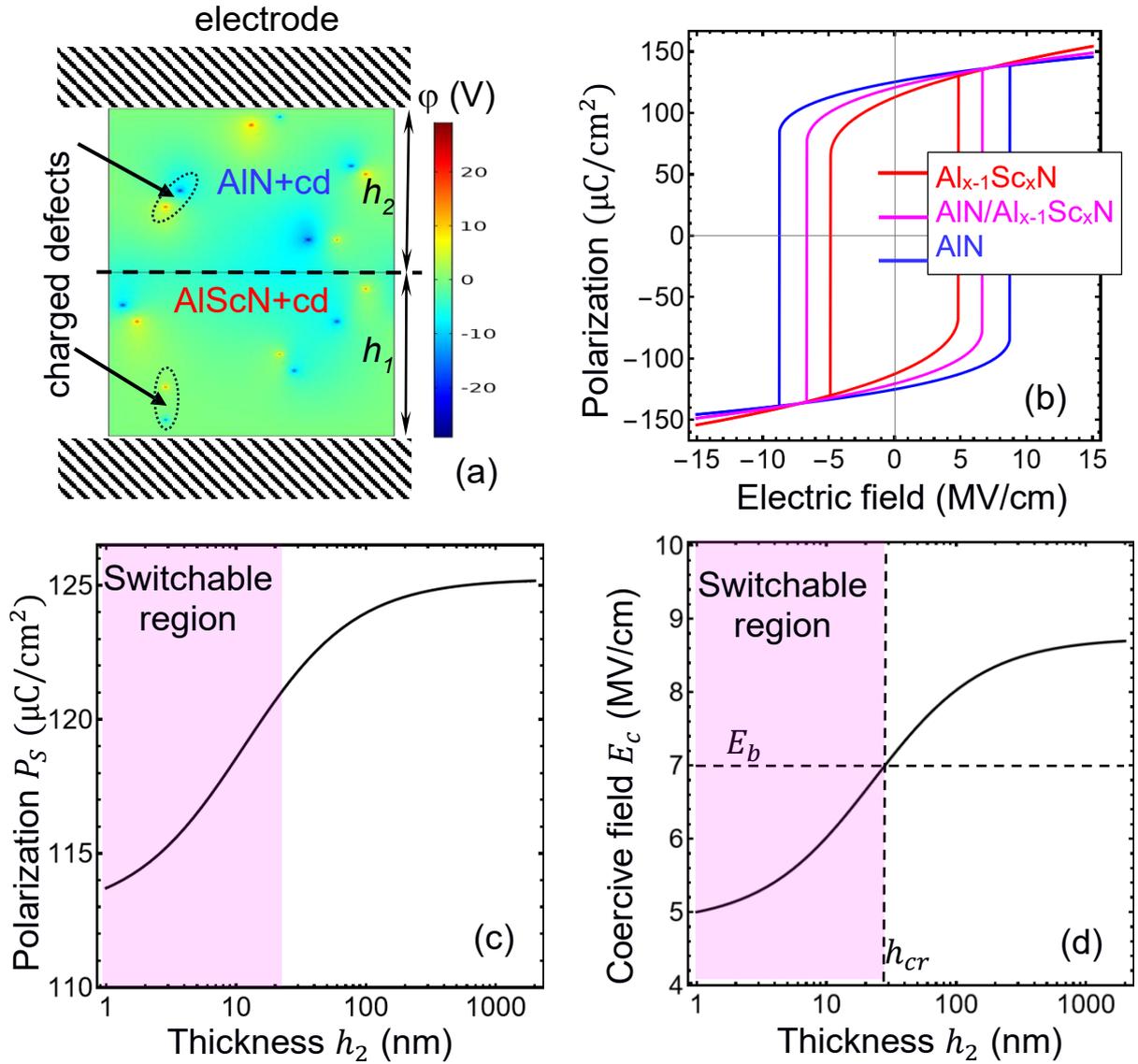

**FIGURE 5.** (a) An illustrative distribution of electrostatic potential in the $AlN/Al_{0.73}Sc_{0.27}N$ bilayer calculated at $U = 0$ in the presence of dipolar-type charged defects (cd), whose parameters are listed in the text. (b) The polarization hysteresis in the $AlN/Al_{0.73}Sc_{0.27}N$ bilayer in the presence of dipolar-type charged defects (the magenta loop, $h_1 = h_2 = 20$ nm). The polarization switching of the Al$_{0.73}$Sc$_{0.27}$N layer (the red loop, $h_1 = 40$ nm $h_2 = 0$) and of the AlN layer (the blue loop, $h_1 = 0$, $h_2 = 40$ nm) calculated in the presence of dipolar-type charged defects disregarding the electric breakdown. The dependence of the bilayer remanent polarization **(c)** and effective coercive field **(d)** on the thickness $h_2$ of AlN layer calculated for the fixed thickness of Al$_{0.73}$Sc$_{0.27}$N layer, $h_1$ =20 nm, in the presence of dipolar-type charged defects. A horizontal dashed line corresponds to the electric breakdown field $E_b$, and its cross-section with the dashed vertical line allows to determine the critical thickness $h_{cr}$ of AlN layer below which the bilayer is switchable by external field. The switchable region is filled with a light-magenta color. The background permittivity $\varepsilon_b^{(1)} = \varepsilon_b^{(2)} = 4$, other LGD parameters are listed in **Table II**.

Since FEM for computational cells thicker than (50 – 100) nm was not possible, the curves in **Figs. 5(c)** and **5(d)** that are calculated for $h_2 \gg 20$ nm are extrapolations for the case when the bilayer polarization behavior depends on the ratio $h_1/h_2$ only. Also, the assumptions about the random distribution of dipolar-type defects in both layers are probably valid for thin layers only. Often defects are accumulated near the surface/interfaces, as well as elastic dipoles, dislocations and/or other topological defects can appear in the layers thicker than (50 – 100) nm. Note, that there are other mechanisms of the coercive field decrease unconsidered in our work, such as topological elastic defects (e.g., dislocations inherent to thicker films) and the strong band-bending induced by the "bare" depolarization field at the moving charged domain wall. Due to the flexoelectric coupling, the dislocation endings can be very strong charged defects [35] leading to the additional coercive field reduction in thicker films. The band-bending can lead to the self-screening of charged domain walls by the free carriers "torn out" by the electric field, being weakly dependent on the layer thickness. The metastable charged domain walls with the saw-tooth shape are observed recently in epitaxial Al$_{1-x}$Sc$_x$N/GaN heterostructures [36]. The redistribution of point charged defects and other types of defects emerging in thicker layers can lead to the weak dependence of the effective coercive field on the thickness of the layers, which is observed experimentally (see e.g., Fig. 4c and 4d and extended experimental data, Fig. 5b and 5d in Ref.[13]).

**Figure 6** illustrates the evolution of polarization distribution in the AlN/Al$_{0.83}$Sc$_{0.27}$N bilayer with $h_1 = h_2$ in the presence of randomly distributed pairs of positive and negative point charges located in the points shown in **Fig. 5(a)**. Each image in **Fig. 6** corresponds to the definite value of the applied field, $E_a = \frac{U}{h_1+h_2}$. It is seen that the random charges, which act as nucleation centers for the domains, are responsible for the appearance and growth of domain needles and filaments between the nearest oppositely charged defects (at $E_a < E_c$), followed by the correlated intergrowth of multiple spike-like domain through the bilayer (at $E_a \approx E_c$), and then to the domain growth in lateral directions (at $E_a > E_c$). The scenario of polarization reversal via the appearance of domain spikes seems common for the difficult-to-switch ferroelectrics and pyroelectrics with a large coercive field, such as LiNbO$_3$ [10], Zn$_{1-x}$Mg$_x$O [7] and particularly for Al$_{x-1}$Sc$_x$N [4].

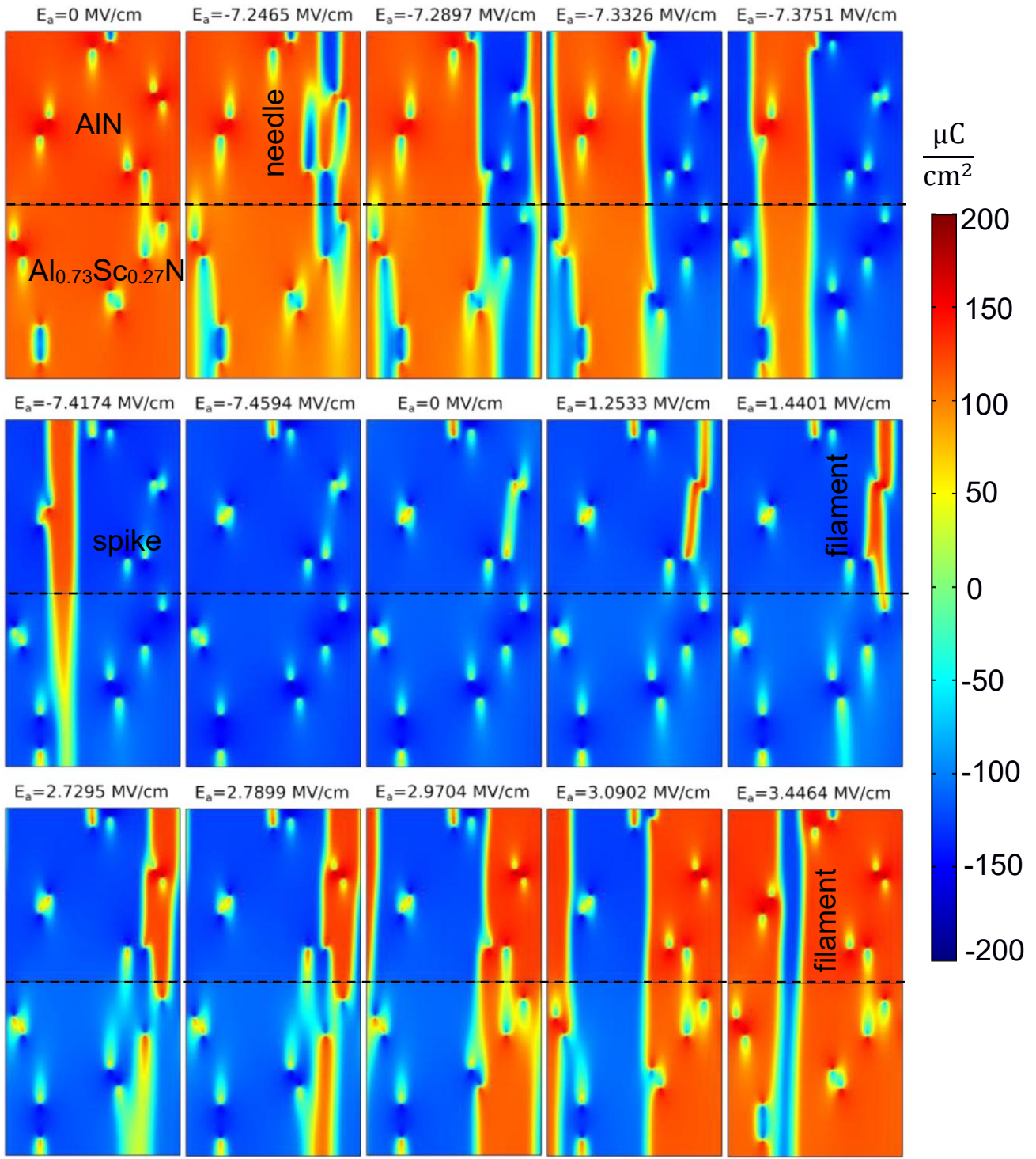

**FIGURE 6**. Evolution of polarization distribution in the AlN/Al$_{0.73}$Sc$_{0.27}$N bilayer with $h_1 = h_2 = 20$ nm. Parameters are the same as in **Fig. 5.** Charged defects in both layers (shown by darker-red or darker-blue small spots) act as nucleation centers.

To summarize the section, the thermodynamic theory of the proximity ferroelectricity, evolved for the *single-domain multilayers*, demonstrates the significant role of the relative thickness of the layers in determining the key properties of the proximity effect, such as spontaneous polarization and effective thermodynamic coercive field. This conclusion becomes invalid in the case when the domain kinetics is considered. In particular, the size effect of coercive field can be much weaker (or virtually absent) due to the defect-induced domain nucleation. Unfortunately, the analytical expressions are available only for the case of the single-domain polarization reversal.

From Figs. 4a-4d in Ref. [13] we can conclude that the coercive fields of four different multilayers, including the bilayer "200 nm AlN / 200 nm AlBN (94:06)" and symmetric three-layer structures "100 nm AlBN (94:06) / 200 nm AlN / 100 nm AlBN (94:06)", "100 nm AlN/ 200 nm AlBN (94:06) / 100 nm AlN" and "75 nm AlBN (94:06) / 500 nm AlN / 75 nm AlBN (94:06)", are close to 6-7 MV/cm and weakly depend on the thickness of AlN layer (200 nm or 500 nm). The thermodynamic theory of proximity ferroelectricity predicts the same thermodynamic coercive field and spontaneous polarization for the multilayers shown in Figs.4a, 4b and 4c in Ref. [13], because the total thickness of AlBN layer(s) is 200 nm and 200 nm for AlN layers. Figure 4d in Ref. [13] shows noticeably higher remanent polarization in comparison with Figs.4a, 4b and 4c therein. This observation also well agrees with the thermodynamic theory of proximity ferroelectricity, because the spontaneous polarization value increases due to the 500-nm thickness of AlN layer (see e.g., **Fig. 5(c)**).

To explain the data shown in Fig. 4d in Ref. [13], we assume that the relative influence of defect-induced domain nucleation is stronger in the "75 nm AlBN (94:06) / 500 nm AlN / 75 nm AlBN (94:06)", which reduces the effective coercive field in the three-layer structure, and the reduced field appears close to one observed in the bilayer structure "200 nm AlBN (94:06) / 200 nm AlN".

While the role of charged defects and charged domain wall dynamics are important to consider for quantitative comparison with experiments, they alone cannot explain the nature of the proximity ferroelectricity per *se*. Defects can *reduce* or *increase* coercive fields [37, 38]. This is where our thermodynamic theory sheds light on the importance of reducing intrinsic switching barriers due to the proximity effect even in the absence of any extrinsic defects as illustrated in Figures 2-4 and Fig. 7 to follow next.

### 5. Proximity-induced ferroelectricity in $Zn_{1-x}Mg_xO$/ZnO bilayers

To demonstrate the generality of the evolved thermodynamical approach, let us consider briefly the proximity effect of the thermodynamic single-domain switching in the $Zn_{1-x}Mg_xO$/ZnO and $Zn_{1-x}Mg_xO$/MgO bilayers with x=0.34. The sketch of the bilayers is shown in **Fig. 7(a). Figure 7(b)** shows the free energy wells of the separate layers, where a shallow red well corresponds to the switchable ferroelectric $Zn_{0.66}Mg_{0.34}O$ (abbreviated as ZMO) bulk material, and a much deeper and wider blue well corresponds to the non-switchable polar ZnO bulk material. LGD parameters of $Zn_{0.66}Mg_{0.34}O$, ZnO and MgO bulk materials, which are listed in **Table III,** were determined from the experimentally measured spontaneous polarization and linear dielectric permittivity [39, 40] as described in **Appendix D** [27]. The background permittivity $\varepsilon_b$ is estimated as the square of refractive index, which is about 1.8 and 1.6 for ZnO and $Zn_{0.66}Mg_{0.34}O$ respectively according to Ref. [41].

**Figure 7(c)** shows the thermodynamic single-domain polarization switching in the $Zn_{0.66}Mg_{0.34}O$/ZnO bilayer capacitor for three thickness of layers, $h_1/h_2$ =0.1, 1 and 10. Let us emphasize that the hysteresis loop corresponding to the single-domain switching of ZnO capacitor cannot be observed experimentally, because the electric breakdown field $E_b$ of the pure ZnO (which is less than 12 MV/cm) is much lower than the calculated intrinsic coercive field, $E_c \approx$ 25 MV/cm (see the green loop in **Fig. 7(c)**). The hysteresis loop corresponding to the single-domain switching of the $Zn_{0.66}Mg_{0.34}O$ capacitor has a high intrinsic coercive field, $E_c \approx$ 7.7 MV/cm (see the red loop in **Fig. 7(c)**). The hysteresis loop corresponding to the single-domain switching of the bilayer $Zn_{0.66}Mg_{0.34}O$/ZnO capacitor, whose coercive field changes from 23 MV/cm (for $h_1/h_2$ =0.1) to 8 MV/cm (for $h_1/h_2$ =10) as shown by the brown loop in **Fig. 7(c)**. Despite the calculated coercive fields being very high, the effect of the ZnO coercivity reduction is caused by the internal electric field, coupling both layers (see Eqs.(5)). Similar to the case of $Al_{x-1}Sc_xN$/AlN capacitors, further reduction of the coercivity in the $Zn_{1-x}Mg_xO$/ZnO capacitors can emerge due to the random charged defects in the bulk of the $Zn_{1-x}Mg_xO$ and/or ZnO layers, which act as nucleation centers for the correlated polarization switching. We defer further investigation of nucleation centers in this material to future studies.

**Figure 7(d)** shows the thermodynamic single-domain polarization switching in the $Zn_{0.66}Mg_{0.34}O$/MgO bilayer capacitor for three thickness of layers, $h_1/h_2$ =1, 10 and 50. Here the proximity of the switchable $Zn_{0.66}Mg_{0.34}O$ layer leads to the collective switching of ferroelectric polarization in both layers for $\frac{h_2}{h_1}$ <0.1 (see **Figs. 7(e)** and **7(f)**). Due to the "anomalous" positive $\alpha$, negative $\beta$ and positive $\gamma$ for $Zn_{0.66}Mg_{0.34}O$, the thin region of double hysteresis loops separates the region of single loops from the dielectric region (see dashed vertical lines in **Fig. 7(f)**). The thickness ratio range $0 < \frac{h_2}{h_1} \leq 0.1$ required for the proximity ferroelectricity in the bilayer is much smaller than the range $0 < \frac{h_2}{h_1} \leq$ (1-2) estimated for $\alpha < 0$, $\beta > 0$ and $\gamma = 0$ (compare **Fig. 7(f)** and **Fig. 2(a)**). Thus, for the observation of the pronounced proximity ferroelectricity in the FE/DE or PE/FE bilayers for the wide range of thickness ratio (e.g., $0 < \frac{h_2}{h_1} < 2$), the second order ferroelectric with $\alpha < 0$ and the dielectric with large dielectric permittivity (e.g., rutile and $SrTiO_3$) are required.

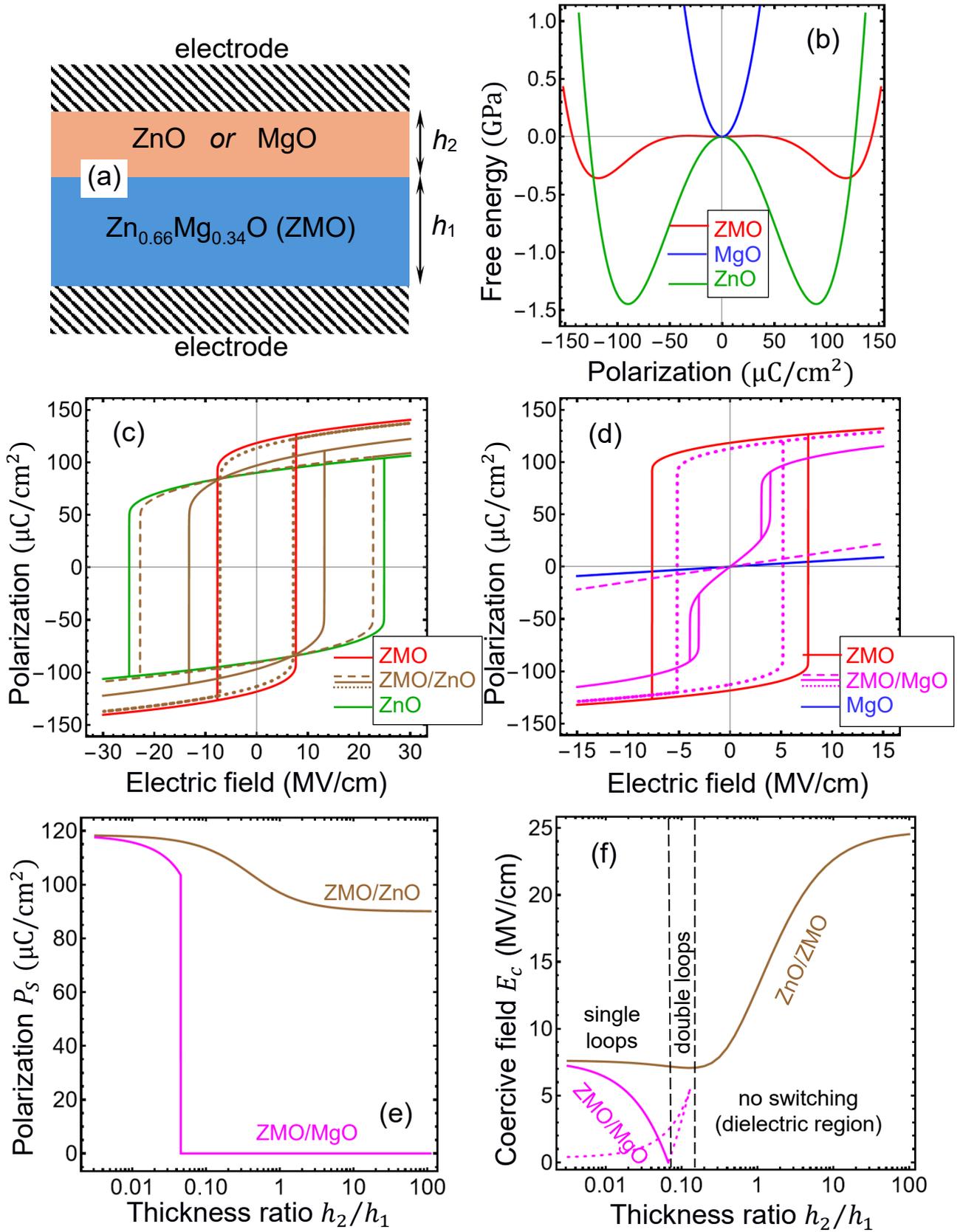

**FIGURE 7. (a)** The sketch of the considered bilayers, $Zn_{0.66}Mg_{0.34}O/ZnO$ (ZMO/ZnO) and $Zn_{0.66}Mg_{0.34}O/MgO$ (ZMO/MgO). **(b)** Free energy wells of the separate layers, where the red, blue and green curves correspond to ZMO, MgO and ZnO, respectively. **(c)** The thermodynamic polarization hysteresis loops calculated for the bulk ZMO (the red loop), "hypothetical" bulk ZnO (the green loop) and ZMO/ZnO bilayers of three thicknesses

ratios, $h_1/h_2 = 0.1$, 1 and 10 (dashed, solid and dotted brown curves, respectively). **(d)** The polarization - field dependencies calculated for the bulk ZMO (the red loop), bulk MgO (the blue curve) and ZMO/MgO bilayers of three thicknesses ratios $h_1/h_2 = 1$, 10 and 50 (dashed, solid and dotted magenta curves, respectively). The dependence of the bilayer remanent polarization **(e)** and effective coercive field **(f)** vs. the thickness ratio $h_2/h_1$ of ZnO layer (brown curves) or MgO layer (magenta curves) to the ZMO layer.

Table III. LGD model parameters for $Zn_{0.66}Mg_{0.34}O$, ZnO and MgO compounds

| compound | $\varepsilon_b$ | $\alpha$, m/F | $\beta$, m$^5$/(F C$^2$) | $\gamma$, m$^7$/(F C$^4$) | property |
|---|---|---|---|---|---|
| $Zn_{0.66}Mg_{0.34}O$ | 2.50 | $2.854 \cdot 10^8$ | $-3.007 \cdot 10^9$ | $2.0 \cdot 10^9$ | switchable |
| ZnO | 3.40 | $-7.152 \cdot 10^9$ | $8.829 \cdot 10^9$ | 0 | non-switchable |
| cubic MgO | 3.02 | $1.659 \cdot 10^{10}$ | 0 | 0 | dielectric |

## 5. Summary

We considered the multilayer film consisting of separate layers with different thickness and different ferroelectric and dielectric properties. Some layers are made of uniaxial ferroelectric materials, whose spontaneous polarization is directed along the multilayer normal vector. Some other layers can be made of dielectric or paraelectric materials, or of non-switchable polar materials. We have shown that the electric polarization of all the layers is strongly coupled due to the long-range internal electric field (which has depolarization and external contributions) in all the abovementioned cases.

The mechanism of the proximity ferroelectricity is an internal electric field determined by the polarization of the layers and their relative thickness in a self-consistent manner (according to Eq.(5)) that renormalizes the double-well ferroelectric potential to lower the steepness of the switching barrier (according to Eq.(6)). Depending on the relative thickness of the layers, the effective coercive field can be significantly lower than the hypothetic thermodynamic coercive fields and/or dielectric breakdown fields of the unswitchable layers. If the collective coercive field appears higher than the dielectric breakdown field, its strong reduction can emerge from the correlated polarization switching in the proximity of random charged defects in the bulk of the layers.

Due to the internal field, either all different-type layers undergo simultaneous ferroelectric switching in a similar way (meaning that hysteresis loops of their individual polarizations have a very similar shape, the same coercive fields and close values of the remanent polarizations), or all layers demonstrate paraelectric or non-switchable behavior simultaneously. We conclude that the simultaneous switching behavior of the layers is the recently discovered [13] proximity ferroelectricity effect, and select the most interesting types of the bilayers, where the polarization switching can reveal the prominent features of the proximity effect. The main features of their proximity-induced polarization switching are summarized below:

The bilayer of the "FE/PE" type, where the first layer of thickness $h_1$ is made of the uniaxial ferroelectric "FE", and the second layer of thickness $h_2$ is made of a paraelectric material "PE". Due to the proximity effect the polarization of both layers undergoes ferroelectric switching for $h_2 \leq h_1$. This happens because the polarization of the paraelectric layer is proportional to the internal electric field; and the field is proportional to the switchable hysteresis-type polarization of the ferroelectric layer.

The bilayer of the "FE/W-FE" and "FE/H-FE" types are where the first layer of thickness $h_1$ is made of the ferroelectric "FE", and the second layer of thickness $h_2$ is made either of a "weak" ferroelectric "W-FE", or of a hard-to-switch ferroelectric material "H-FE" with a very large coercive field (but still below the electric breakdown field). Due to the proximity effect, the polarization in both layers undergoes simultaneous ferroelectric switching in a very similar manner. Namely, the remanent values of the layers' polarizations are very close, $P_s^{(1)} \approx P_s^{(2)} \approx P_s$, the coercive fields are exactly the same, $E_c^{(1)} = E_c^{(2)} = E_c$, and loop shapes are very similar, independent of the thickness ratio $\frac{h_1}{h_2}$. However, the magnitudes of $P_s$ and $E_c$ are determined by the ratio $\frac{h_1}{h_2}$.

The bilayer of the "FE/N-FE" type, where the first layer of thickness $h_1$ is made of the ferroelectric "FE" and the second layer of thickness $h_2$ is made of the non-switchable polar material "N-FE" with a very high thermodynamic coercive field, being well above the electric breakdown field. The dielectric permittivity, $\varepsilon_2$, of the second layer is regarded to be much smaller than the permittivity, $\varepsilon_1$, of the switchable layer. The changes in the applied electric field, induced in both layers by the mismatch of their dielectric permittivity ($\varepsilon_1 \gg \varepsilon_2$), tends to reduce the apparent internal field in the N-FE layer due to the splitting of external field inside the layers, which can be roughly estimated as $\frac{E_z^{(2)}}{E_z^{(1)}} \sim \frac{\varepsilon_1}{\varepsilon_2} \gg 1$. As a result, we observe the significant reduction in the effective coercive field $E_c$. Due to the proximity effect (which consists in the depolarization field appearance and applied field splitting) both layers undergo the simultaneous ferroelectric switching in a very similar manner: the hysteresis loops of polarizations $P_z^{(1)}$ and $P_z^{(2)}$ have the same coercive fields and remanent polarizations. This formally means that the reduction in the effective coercive field of the second N-FE layer is due to the changes in its double-well potential. However, the physical origin of the potential change is the layers coupling by the internal electric fields. Due to the proximity effect, the polarization of both layers undergoes ferroelectric switching for $h_2 \leq h_{cr}$, where the critical thickness is determined by the electric breakdown field of the bilayer (often $h_{cr} \geq h_1$).

The above proposed mechanism of the "FE/N-FE" bilayer switching was used to describe quantitatively the experimentally observed proximity-induced polarization switching in Al$_{1-x}$Sc$_x$N/AlN and Mg$_x$Zn$_{1-x}$O/ZnO bilayers. Additional reduction of the coercive field to the values observed experimentally in $Al_{1-x}Sc_xN$ (below the electric breakdown field) is reached due to the random electric charges, which act as nucleation centers for the domains. The charges are responsible for the appearance and growth of domain needles and filaments between the defects (below the coercive fields), followed by the correlated intergrowth of multiple spike-like domains through the bilayer (near the coercive field), and then to the expansion of the domains in the lateral direction (above the coercive field).

The proposed theory goes beyond understanding the current experiments to predict new phenomena involving dielectric stacks of ferroelectrics with paraelectrics and dielectrics, rendering the latter into switchable ferroelectrics as well. This opens up exciting avenues for future experiments, opening up a vastly larger universe of ferroelectrics. The proposed proximity-induced polarization switching will be important in the material design of Si-compatible ferroelectric bilayers, such as $ZrO_2/Hf_xZr_{1-x}O_2$, $ZnO/Zn_{1-x}Mg_xO$ and $AlN/Al_{1-x}Sc_xN$, for advanced nanoelectronics and optoelectronics.

**Acknowledgements.** Authors are very grateful to the Referees for useful suggestions and stimulating discussions. The work is supported by the DOE Software Project on "Computational Mesoscale Science and Open Software for Quantum Materials", under Award Number DE-SC0020145 as part of the Computational Materials Sciences Program of US Department of Energy, Office of Science, Basic Energy Sciences. Results were visualized in Mathematica 14.0 [42] (see https://notebookarchive.org/2025-01-de3c6bf). V.G. and J-P.M. acknowledges support from the Center for 3D Ferroelectric Microelectronics (3DFeM), an Energy Frontier Research Center funded by the U.S. Department of Energy, Office of Science, Office of Basic Energy Sciences Energy Frontier Research Centers program under Award Number DE-SC0021118. The experimental results that motivated this theoretical study were generated and shared by J-P.M.

**Authors' contribution.** A.N.M. and V.G. generated the research idea, formulated the problem and wrote the manuscript draft. E.A.E. wrote the codes, performed analytical calculations and prepared figures jointly with A.N.M. A.N.M., L.-Q.C. and V.G. worked on the results explanation and manuscript improvement. All co-authors discussed the results.

# Supplementary Materials to the Manuscript
# "A Thermodynamic Theory of Proximity Ferroelectricity"

## Appendix A. Approximate static solutions for the single-domain bilayer

The coupled equations for the layer polarizations are

$$\alpha_1 P_z^{(1)} + \beta_1 \left(P_z^{(1)}\right)^3 + \gamma_1 \left(P_z^{(1)}\right)^5 = E_z^{(1)}, \tag{A.1a}$$

$$\alpha_2 P_z^{(2)} + \beta_2 \left(P_z^{(2)}\right)^3 + \gamma_2 \left(P_z^{(2)}\right)^5 = E_z^{(2)}. \tag{A.1b}$$

Hereinafter we suppose that the single-domain polarization is homogeneous inside each layer (due to the natural boundary conditions). $\alpha_i$, $\beta_i$ and $\gamma_i$ are the LGD expansion coefficients of the individual layers; $E_z^{(i)}$ is the electric field inside layer "$i$". In the considered case

$$E_z^{(i)} = -\frac{P_z^{(i)} - \bar{D}}{\varepsilon_0 \varepsilon_b^{(i)}} + \frac{U}{\varepsilon_b^{(i)} \left[h_1/\varepsilon_b^{(1)} + h_2/\varepsilon_b^{(2)}\right]}, \tag{A.2}$$

where $\varepsilon_b^{(i)}$ is the background permittivity of the layer "$i$", $\bar{D}$ is the average displacement of the bilayer:

$$\bar{D} = \frac{\frac{h_1}{\varepsilon_b^{(1)}} P_z^{(1)} + \frac{h_2}{\varepsilon_b^{(2)}} P_z^{(2)}}{\frac{h_1}{\varepsilon_b^{(1)}} + \frac{h_2}{\varepsilon_b^{(2)}}}. \tag{A.3}$$

Using the expressions for the internal field (A.2) and average polarization (A.3), one could rewrite Eqs.(A.1) as follows

$$(\alpha_1 + \aleph_1) P_z^{(1)} + \beta_1 \left(P_z^{(1)}\right)^3 + \gamma_1 \left(P_z^{(1)}\right)^5 - \aleph_1 P_z^{(2)} = \eta_1 E_a. \tag{A.4a}$$

$$(\alpha_2 + \aleph_2) P_z^{(2)} + \beta_2 \left(P_z^{(2)}\right)^3 + \gamma_2 \left(P_z^{(2)}\right)^5 - \aleph_2 P_z^{(1)} = \eta_2 E_a. \tag{A.4b}$$

In Eqs.(A.4) we introduced the depolarization field factors

$$\aleph_1 \stackrel{\text{def}}{=} \frac{h_2/\varepsilon_b^{(2)}}{\varepsilon_0 \varepsilon_b^{(1)} \left[h_1/\varepsilon_b^{(1)} + h_2/\varepsilon_b^{(2)}\right]}, \quad \aleph_2 \stackrel{\text{def}}{=} \frac{h_1/\varepsilon_b^{(1)}}{\varepsilon_0 \varepsilon_b^{(2)} \left[h_1/\varepsilon_b^{(1)} + h_2/\varepsilon_b^{(2)}\right]}, \tag{A.5a}$$

the effective field factors

$$\eta_1 \stackrel{\text{def}}{=} \frac{h_1 + h_2}{\varepsilon_b^{(1)} \left[h_1/\varepsilon_b^{(1)} + h_2/\varepsilon_b^{(2)}\right]}, \quad \eta_2 \stackrel{\text{def}}{=} \frac{h_1 + h_2}{\varepsilon_b^{(2)} \left[h_1/\varepsilon_b^{(1)} + h_2/\varepsilon_b^{(2)}\right]}, \tag{A.5b}$$

and the average electric field

$$E_a \stackrel{\text{def}}{=} \frac{U}{h_1 + h_2}. \tag{A.5c}$$

Since $|\alpha_i| \varepsilon_0 \varepsilon_b^{(i)} \ll 1$ for most ferroelectrics, the following strong inequalities are valid

$$|\alpha_i| \ll \aleph_i \tag{A.5d}$$

The only exceptions are the cases of extremely different values of thickness of the layers.

One could rewrite Eq.(A.4a) as $P_z^{(2)} = \left\{(\alpha_1 + \aleph_1) P_z^{(1)} + \beta_1 \left(P_z^{(1)}\right)^3 + \gamma_1 \left(P_z^{(1)}\right)^5 - \eta_1 E_a\right\}/\aleph_1$ and substitute it into Eq.(A.4b) to get the equation for $P_1$:

$$(\alpha_2 + \aleph_2)\left[\frac{(\alpha_1+\aleph_1)P_z^{(1)}+\beta_1\left(P_z^{(1)}\right)^3+\gamma_1\left(P_z^{(1)}\right)^5-\eta_1 E_a}{\aleph_1}\right] - \aleph_2 P_1 + \beta_2\left[\frac{(\alpha_1+\aleph_1)P_z^{(1)}+\beta_1\left(P_z^{(1)}\right)^3+\gamma_1\left(P_z^{(1)}\right)^5-\eta_1 E_a}{\aleph_1}\right]^3 +$$

$$\gamma_2\left[\frac{(\alpha_1+\aleph_1)P_z^{(1)}+\beta_1\left(P_z^{(1)}\right)^3+\gamma_1\left(P_z^{(1)}\right)^5-\eta_1 E_a}{\aleph_1}\right]^5 = \eta_2 E_a \qquad (A.6a)$$

It is seen that many terms in the higher-order Eq.(A.6a) are proportional to the small parameters like $\alpha_1/\aleph_1$. Therefore, neglecting the higher-order terms with respect to the polarization, one could yield for following equation for the polarization $P_z^{(1)}$ at zero bias field ($E_a = 0$):

$$P_z^{(1)}\left(\alpha_2 + \frac{\alpha_1(\alpha_2+\aleph_2)}{\aleph_1}\right) + \left(P_z^{(1)}\right)^3\left(\beta_2\left\{1+\frac{\alpha_1}{\aleph_1}\right\}^3 + \frac{\beta_1(\alpha_2+\aleph_2)}{\aleph_1}\right) + \left(P_z^{(1)}\right)^5\left(\gamma_2\left\{1+\frac{\alpha_1}{\aleph_1}\right\}^5 + \frac{\gamma_1(\alpha_2+\aleph_2)}{\aleph_1} + \right.$$

$$\left.\left\{1+\frac{\alpha_1}{\aleph_1}\right\}^2\frac{3\beta_1\beta_2}{\aleph_1}\right) + \frac{\left(P_z^{(1)}\right)^7}{\aleph_1}\left\{1+\frac{\alpha_1}{\aleph_1}\right\}\left(\frac{3\beta_1^2\beta_2}{\aleph_1} + 3\beta_2\gamma_1\left\{1+\frac{\alpha_1}{\aleph_1}\right\} + 5\beta_1\gamma_2\left\{1+\frac{\alpha_1}{\aleph_1}\right\}^3\right) + \cdots = 0 \qquad (A.6b)$$

Here one could neglect all the terms with powers higher than the fifth, since they have coefficients inversely proportional to the powers of $\aleph_1$. In a similar way one could obtain from Eqs.(A.4) the equation for $P_z^{(2)}$:

$$(\alpha_1 + \aleph_1)\left[\frac{(\alpha_2+\aleph_2)P_z^{(2)}+\beta_2\left(P_z^{(2)}\right)^3+\gamma_2\left(P_z^{(2)}\right)^5-\eta_2 E_a}{\aleph_2}\right] + \beta_1\left[\frac{(\alpha_2+\aleph_2)P_z^{(2)}+\beta_2\left(P_z^{(2)}\right)^3+\gamma_2\left(P_z^{(2)}\right)^5-\eta_2 E_a}{\aleph_2}\right]^3 +$$

$$\gamma_1\left[\frac{(\alpha_2+\aleph_2)P_z^{(2)}+\beta_2\left(P_z^{(2)}\right)^3+\gamma_2\left(P_z^{(2)}\right)^5-\eta_2 E_a}{\aleph_2}\right]^5 - \aleph_1 P_z^{(2)} = \eta_1 E_a \qquad (A.7)$$

Based on Eqs.(A.6b) and (A.7), one could derive the equations for the polarization of the different layers (at small external field) as follows:

$$\alpha_1^R P_z^{(1)} + \beta_1^{(R)}\left(P_z^{(1)}\right)^3 + \gamma_1^{(R)}\left(P_z^{(1)}\right)^5 \approx \eta_1^{(R)} E_a, \qquad (A.8a)$$

$$\alpha_2^R P_z^{(2)} + \beta_2^{(R)}\left(P_z^{(2)}\right)^3 + \gamma_2^{(R)}\left(P_z^{(2)}\right)^5 \approx \eta_2^{(R)} E_a. \qquad (A.8b)$$

Here we introduced the renormalized coefficients

$$\alpha_1^R \stackrel{\text{def}}{=} \frac{\alpha_2\aleph_1+\alpha_1\alpha_2+\alpha_1\aleph_2}{\aleph_1}, \quad \alpha_2^R \stackrel{\text{def}}{=} \frac{\alpha_2\aleph_1+\alpha_1\alpha_2+\alpha_1\aleph_2}{\aleph_2}, \qquad (A.9a)$$

$$\beta_1^{(R)} \stackrel{\text{def}}{=} \beta_2\left\{1+\frac{\alpha_1}{\aleph_1}\right\}^3 + \frac{\beta_1(\alpha_2+\aleph_2)}{\aleph_1}, \quad \beta_2^{(R)} \stackrel{\text{def}}{=} \beta_1\left\{1+\frac{\alpha_2}{\aleph_2}\right\}^3 + \frac{\beta_2(\alpha_1+\aleph_1)}{\aleph_2}, \qquad (A.9b)$$

$$\gamma_1^{(R)} \stackrel{\text{def}}{=} \gamma_2\left\{1+\frac{\alpha_1}{\aleph_1}\right\}^5 + \frac{\gamma_1(\alpha_2+\aleph_2)}{\aleph_1} + \left\{1+\frac{\alpha_1}{\aleph_1}\right\}^2\frac{3\beta_1\beta_2}{\aleph_1}, \qquad (A.9c)$$

$$\gamma_2^{(R)} \stackrel{\text{def}}{=} \gamma_1\left\{1+\frac{\alpha_2}{\aleph_2}\right\}^5 + \frac{\gamma_2(\alpha_1+\aleph_1)}{\aleph_2} + \left\{1+\frac{\alpha_2}{\aleph_2}\right\}^2\frac{3\beta_1\beta_2}{\aleph_2} \qquad (A.9d)$$

$$\eta_1^{(R)} \stackrel{\text{def}}{=} \eta_2 + \frac{\eta_1}{\aleph_1}\left(\alpha_2 + \aleph_2 + 3\beta_2\left[\frac{(\alpha_1+\aleph_1)P_z^{(1)}+\beta_1\left(P_z^{(1)}\right)^3+\gamma_1\left(P_z^{(1)}\right)^5}{\aleph_1}\right]^2 + \right.$$

$$\left. 5\gamma_2\left[\frac{(\alpha_1+\aleph_1)P_z^{(1)}+\beta_1\left(P_z^{(1)}\right)^3+\gamma_1\left(P_z^{(1)}\right)^5}{\aleph_1}\right]^4\right) \quad \text{(A.9e)}$$

$$\eta_2^{(R)} \stackrel{\text{def}}{=} \eta_1 + \frac{\eta_2}{\aleph_2}\left(\alpha_1 + \aleph_1 + 3\beta_1\left[\frac{(\alpha_2+\aleph_2)P_z^{(2)}+\beta_2\left(P_z^{(2)}\right)^3+\gamma_2\left(P_z^{(2)}\right)^5}{\aleph_2}\right]^2 + \right.$$

$$\left. 5\gamma_1\left[\frac{(\alpha_2+\aleph_2)P_z^{(2)}+\beta_2\left(P_z^{(2)}\right)^3+\gamma_2\left(P_z^{(2)}\right)^5}{\aleph_2}\right]^4\right). \quad \text{(A.9f)}$$

The spontaneous polarization can be found from Eqs.(A.8) and (A.9) as

$$P_{is} = \sqrt{\frac{-2\alpha_i^{(R)}}{\beta_i^{(R)}+\sqrt{\left(\beta_i^{(R)}\right)^2-4\alpha_i^{(R)}\gamma_i^{(R)}}}} \quad \text{for layers } i=1, 2. \quad \text{(A.10)}$$

The coercive field of the bilayer polarization hysteresis could be derived as follows. It is well known that not only the polarization changes its sign at the reversion point, but also the derivative of the polarization with respect to the electric field diverges. Differentiation of Eqs.(A.4) with respect to $E_a$ gives the following equation

$$\left(\alpha_1 + \aleph_1 + 3\beta_1\left(P_z^{(1)}\right)^2 + 5\gamma_1\left(P_z^{(1)}\right)^4\right)\frac{\partial P_z^{(1)}}{\partial E_a} - \aleph_1 \frac{\partial P_z^{(2)}}{\partial E_a} = \eta_1, \quad \text{(A.11a)}$$

$$\left(\alpha_2 + \aleph_2 + 3\beta_2\left(P_z^{(2)}\right)^2 + 5\gamma_2\left(P_z^{(2)}\right)^4\right)\frac{\partial P_z^{(2)}}{\partial E_a} - \aleph_2 \frac{\partial P_z^{(1)}}{\partial E_a} = \eta_2. \quad \text{(A.11b)}$$

The solution of the system of Eqs.(A.11) diverges only under the condition of zero determinant. After excluding the electric field from the equations (A.4) one could get the following system of equations determining the polarization of the layers at the coercive field

$$\begin{cases}\left(\alpha_1 + \aleph_1 + 3\beta_1\left(P_z^{(1)}\right)^2 + 5\gamma_1\left(P_z^{(1)}\right)^4\right)\left(\alpha_2 + \aleph_2 + 3\beta_2\left(P_z^{(2)}\right)^2 + 5\gamma_2\left(P_z^{(2)}\right)^4\right) - \aleph_1\aleph_2 = 0, \\ \frac{(\alpha_1+\aleph_1)P_z^{(1)}+\beta_1\left(P_z^{(1)}\right)^3+\gamma_1\left(P_z^{(1)}\right)^5-\aleph_1 P_2}{\eta_1} = \frac{(\alpha_2+\aleph_2)P_z^{(2)}+\beta_2\left(P_z^{(2)}\right)^3+\gamma_2\left(P_z^{(2)}\right)^5-\aleph_2 P_z^{(1)}}{\eta_2}.\end{cases}$$

(A.12)

For the common case $\frac{|\alpha_i|}{\aleph_i} \ll 1$ and arbitrary ratio $\frac{\aleph_1}{\aleph_2} = \frac{h_2}{h_1}$, we obtained from Eqs.(A.9) and Eqs.(6) (in the main text) that $\alpha_1^{(R)} \approx \frac{\alpha_2\aleph_1+\alpha_1\alpha_2+\alpha_1\aleph_2}{\aleph_1}$, $\beta_1^{(R)} \approx \frac{\beta_2\aleph_1+\beta_1\aleph_2}{\aleph_1}$, $\gamma_1^{(R)} \approx \frac{\gamma_2\aleph_1+\gamma_1\aleph_2+3\beta_1\beta_2}{\aleph_1}$, $\alpha_2^{(R)} \approx \frac{\alpha_2\aleph_1+\alpha_1\alpha_2+\alpha_1\aleph_2}{\aleph_2}$, $\beta_2^{(R)} \approx \frac{\beta_1\aleph_2+\beta_2\aleph_1}{\aleph_2}$, $\gamma_2^{(R)} \approx \frac{\gamma_1\aleph_2+\gamma_2\aleph_1+3\beta_1\beta_2}{\aleph_2}$, $\eta_1^{(R)} \approx \eta_2 + \frac{\aleph_2}{\aleph_1}\eta_1$ and $\eta_2^{(R)} \approx \eta_1 + \frac{\aleph_1}{\aleph_2}\eta_2$. Thus the ratios of the corresponding renormalized coefficients are the almost same, namely $\frac{\alpha_1^{(R)}}{\alpha_2^{(R)}} \approx \frac{\beta_1^{(R)}}{\beta_2^{(R)}} \approx \frac{\gamma_1^{(R)}}{\gamma_2^{(R)}} \approx \frac{\eta_1^{(R)}}{\eta_2^{(R)}} \approx \frac{\aleph_2}{\aleph_1} \approx \frac{h_1}{h_2}$. This means that the remanent polarizations of the layers, which are given by expressions

$$P_s^{(i)} = \sqrt{\frac{-2\alpha_i^{(R)}}{\beta_i^{(R)} + \sqrt{\left(\beta_i^{(R)}\right)^2 - 4\alpha_i^{(R)}\gamma_i^{(R)}}}}, \qquad (A.13a)$$

are almost the same with an accuracy of the ratio $\frac{|\alpha_i|}{\aleph_i}$. Notably, that thermodynamic coercive fields, which are given by expressions

$$E_c^{(i)} = \frac{2}{5}\left(2\beta_i^{(R)} + \sqrt{\left(3\beta_i^{(R)}\right)^2 - 20\alpha_i^{(R)}\gamma_i^{(R)}}\right)\left(\frac{2\alpha_i^{(R)}}{-3\beta_i^{(R)} - \sqrt{\left(3\beta_i^{(R)}\right)^2 - 20\alpha_i^{(R)}\gamma_i^{(R)}}}\right)^{3/2}, \qquad (A.13b)$$

are different and related as $\frac{E_c^{(1)}}{E_c^{(2)}} \approx \frac{\aleph_2}{\aleph_1} \approx \frac{h_1}{h_2}$. The "effective" coercive fields, $E_{ca}^{(i)} = \frac{E_c^{(i)}}{\eta_i^{(R)}}$, which are included in the right-hand sides of Eqs.(A.8), are given by expressions

$$E_{ca}^{(i)} = \frac{2}{5\eta_i^{(R)}}\left(2\beta_i^{(R)} + \sqrt{\left(3\beta_i^{(R)}\right)^2 - 20\alpha_i^{(R)}\gamma_i^{(R)}}\right)\left(\frac{2\alpha_i^{(R)}}{-3\beta_i^{(R)} - \sqrt{\left(3\beta_i^{(R)}\right)^2 - 20\alpha_i^{(R)}\gamma_i^{(R)}}}\right)^{3/2}. \qquad (A.13c)$$

They are related as $\frac{E_{ca}^{(1)}}{E_{ca}^{(2)}} \approx \frac{\eta_2^{(R)}}{\eta_1^{(R)}}\frac{\aleph_2}{\aleph_1} \approx 1$, and so $E_{ca}^{(1)} \approx E_{ca}^{(2)}$.

The relation $E_{ca}^{(1)} \approx E_{ca}^{(2)}$ could be understood from the following back-envelope consideration. The splitting of the electric field, applied to the bilayer capacitor, is proportional to the layers' dielectric permittivity ratio, $\frac{E_a^{(1)}}{E_a^{(2)}} = \frac{\varepsilon_2}{\varepsilon_1}$. The applied electric fields existing in the layers can be found from the system of equations $\frac{E_a^{(1)}}{E_a^{(2)}} = \frac{\varepsilon_2}{\varepsilon_1}$ and $E_a^{(1)} h_1 + E_a^{(2)} h_2 = U$, which gives $E_a^{(1)} = \frac{\varepsilon_2 U}{\varepsilon_2 h_1 + \varepsilon_1 h_2}$ and $E_a^{(2)} = \frac{\varepsilon_1 U}{\varepsilon_2 h_1 + \varepsilon_1 h_2}$. The "effective" coercive voltages can be estimated from the condition $E_a^{(i)} = E_c^{(i)}$ (where $E_c^{(i)}$ is given by Eqs.(A.13b)), which leads to expressions

$$U_{c1} = \frac{\varepsilon_2 h_1 + \varepsilon_1 h_2}{\varepsilon_2} E_c^{(1)}, \qquad U_{c2} = \frac{\varepsilon_2 h_1 + \varepsilon_1 h_2}{\varepsilon_1} E_c^{(2)}. \qquad (A.14)$$

Let us estimate the ratio

$$\frac{\varepsilon_2 - \varepsilon_b^{(2)}}{\varepsilon_1 - \varepsilon_b^{(1)}} \approx \frac{\alpha_1^{(R)} + 3\beta_1^{(R)}\left(P_s^{(1)}\right)^2 + 5\gamma_1^{(R)}\left(P_s^{(1)}\right)^4}{\alpha_2^{(R)} + 3\beta_2^{(R)}\left(P_s^{(2)}\right)^2 + 5\gamma_2^{(R)}\left(P_s^{(2)}\right)^4} \approx \frac{\aleph_2}{\aleph_1} \approx \frac{h_1}{h_2}. \qquad (A.15)$$

Thus, if the background permittivity $\varepsilon_b^{(i)}$ is much smaller than the total permittivity $\varepsilon_i$, we obtain that $\frac{\varepsilon_2}{\varepsilon_1} \approx \frac{h_1}{h_2}$. Under the conditions $\frac{\varepsilon_2}{\varepsilon_1} \approx \frac{h_1}{h_2}$ and $\frac{E_c^{(1)}}{E_c^{(2)}} \approx \frac{h_1}{h_2}$ we obtain from Eq.(A.14) that $U_{c1} \approx U_{c2} \approx U_c$. This means that the coercive field is the same in both layers being equal to

$$E_{ca}^{(i)} \approx \frac{U_c}{h_1+h_2}. \tag{A.16}$$

**Figure A1** demonstrates the effect of collective switching in the bilayer, when the layers collectively switch in the regime of "proximity switching", as well as the regime of "proximity suppression", where they collectively do not switch.

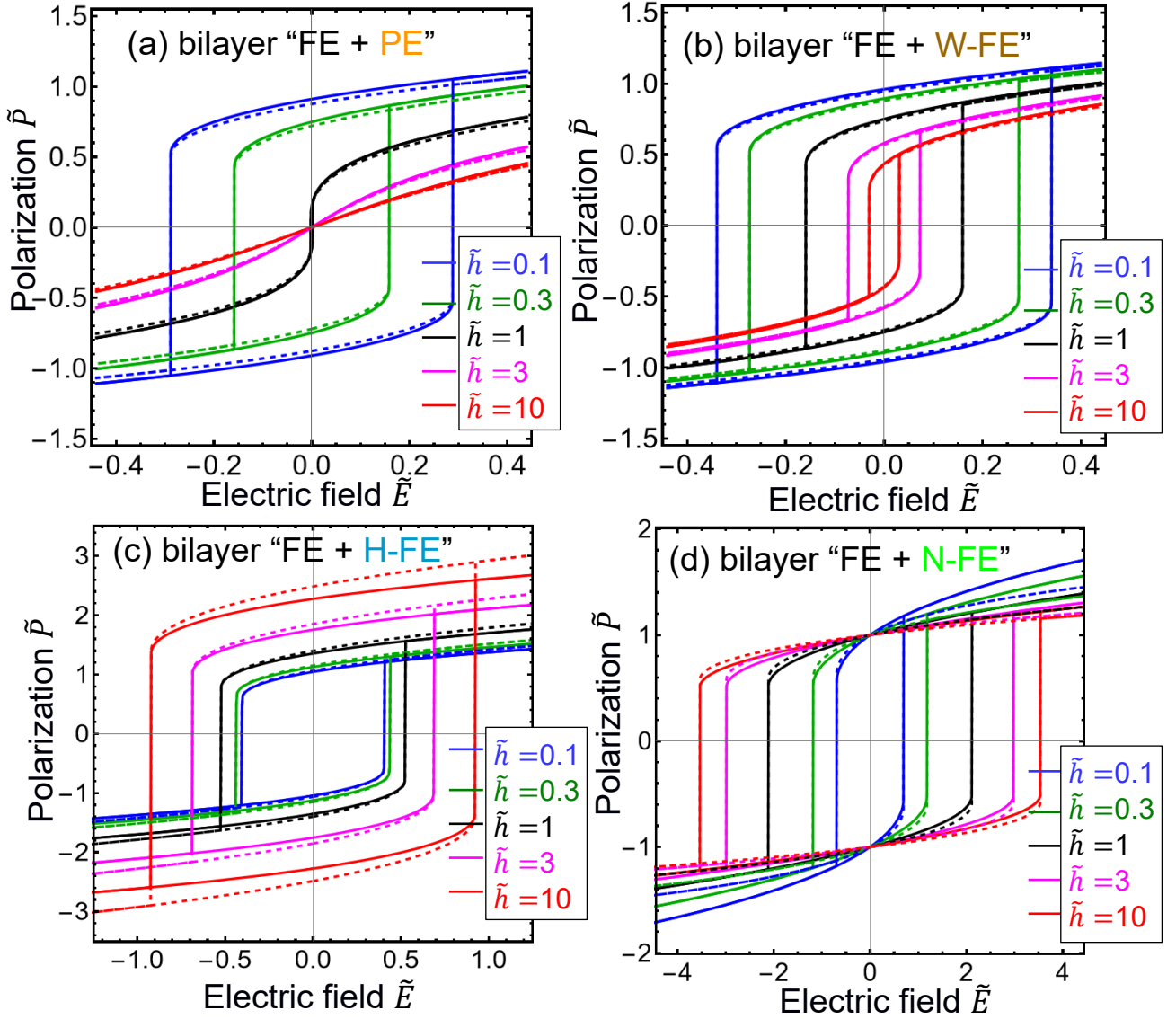

**FIGURE A1**. Hysteresis loops of bilayer polarizations $\tilde{P}_z^{(1)}(\tilde{E})$ (solid curves) and $\tilde{P}_z^{(2)}(\tilde{E})$ (dashed curves) calculated for the fixed coefficients of the first layer ($\tilde{\alpha} = -1, \tilde{\beta} = 1$) and different coefficients of the second layer $\tilde{\alpha} = +1, \tilde{\beta} = 1$ (**a**), $\tilde{\alpha} = -0.1, \tilde{\beta} = 1$ (**b**), $\tilde{\alpha} = -1, \tilde{\beta} = 0.1$ (**c**), and $\tilde{\alpha} = -10, \tilde{\beta} = 10$ (**d**). The thickness ratio $\tilde{h}$=0.1, 0.3, 1, 3 and 10 (blue, green, black, magenta and red curves, respectively). Other parameters are $\tilde{\gamma}_1 = \tilde{\gamma}_2 = 0$ and $\xi_1 = \xi_2 = 0.02$.

**Appendix B. The dependences of the remanent polarization and coercive field on the LGD parameters**

The dependences of the dimensionless remanent polarization $\tilde{P}_s$ and intrinsic coercive field $\tilde{E}_c$ of bilayer hysteresis loops on the dimensionless LGD coefficients $\tilde{\alpha}$ and $\tilde{\beta}$, are shown in **Fig. B1** for different values of the layers thickness ratio $\tilde{h}$ (namely $10 \leq \tilde{h} \leq 0.1$). The almost vertical boundary between the FE and PE phases means the actual independence of the FE-PE transition on the $\tilde{\beta}$ value. The $\tilde{\alpha}$-position of the FE-PE boundary strongly shifts to the left with increase in $\tilde{h}$, because positive and large $\tilde{\alpha}$ favors the PE state of the layer 2, while the layer 1 is made of a ferroelectric material ($\alpha_1 < 0$). Notably that pronounced hysteresis loops can be observed in the case $\tilde{h} \gg 1$, which is a manifestation of the proximity-induced ferroelectric switching effect.

The dependences of the dimensionless remanent polarization $\tilde{P}_s$ and intrinsic coercive field $\tilde{E}_c$ of bilayer hysteresis loops on the dimensionless LGD coefficients $\tilde{\alpha}$ and thickness ratio $\tilde{h}$ are shown in **Fig. B2** for different values of the LGD coefficient $\tilde{\beta}$ (namely $10 \leq \tilde{\beta} \leq 0.1$). The boundary between the FE and PE phases are curved meaning the relatively strong dependence of the transition conditions on the $\tilde{h}$ value. The $\tilde{\alpha}$-position of the FE-PE boundary is almost independent on the $\tilde{\beta}$ value, but the contrast in the FE phase depend on the $\tilde{\beta}$ value, because positive and big $\tilde{\beta}$ favors the weak ferroelectric properties of the layer 2 in comparison with the layer 1. Notably that pronounced hysteresis loops can be observed in the case $\tilde{h} \gg 1$, which is a manifestation of the proximity-induced ferroelectric switching effect.

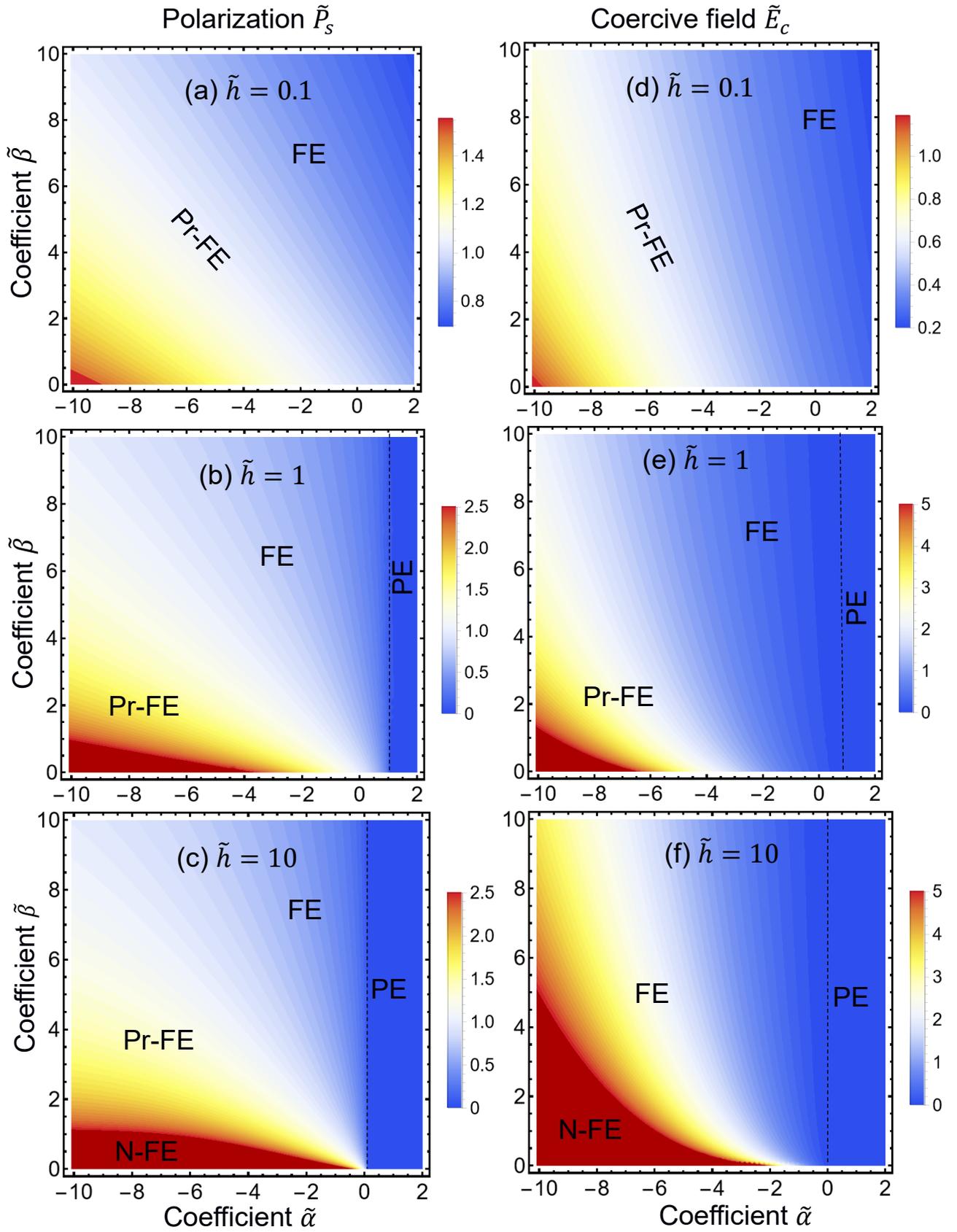

**FIGURE B1**. The dependences of the dimensionless remanent polarization $\tilde{P}_s$ **(a, b, c)** and coercive field $\tilde{E}_c$ **(d, e, f)** of the bilayer polarization hysteresis loops on the dimensionless LGD coefficients $\tilde{\alpha}$ and $\tilde{\beta}$ for different values of the layers thickness ratio $\tilde{h}$=0.1 **(a, d)**, 1 **(b, e)** and 10 **(c, f)**. Other parameters are the same as in **Fig. 2**.

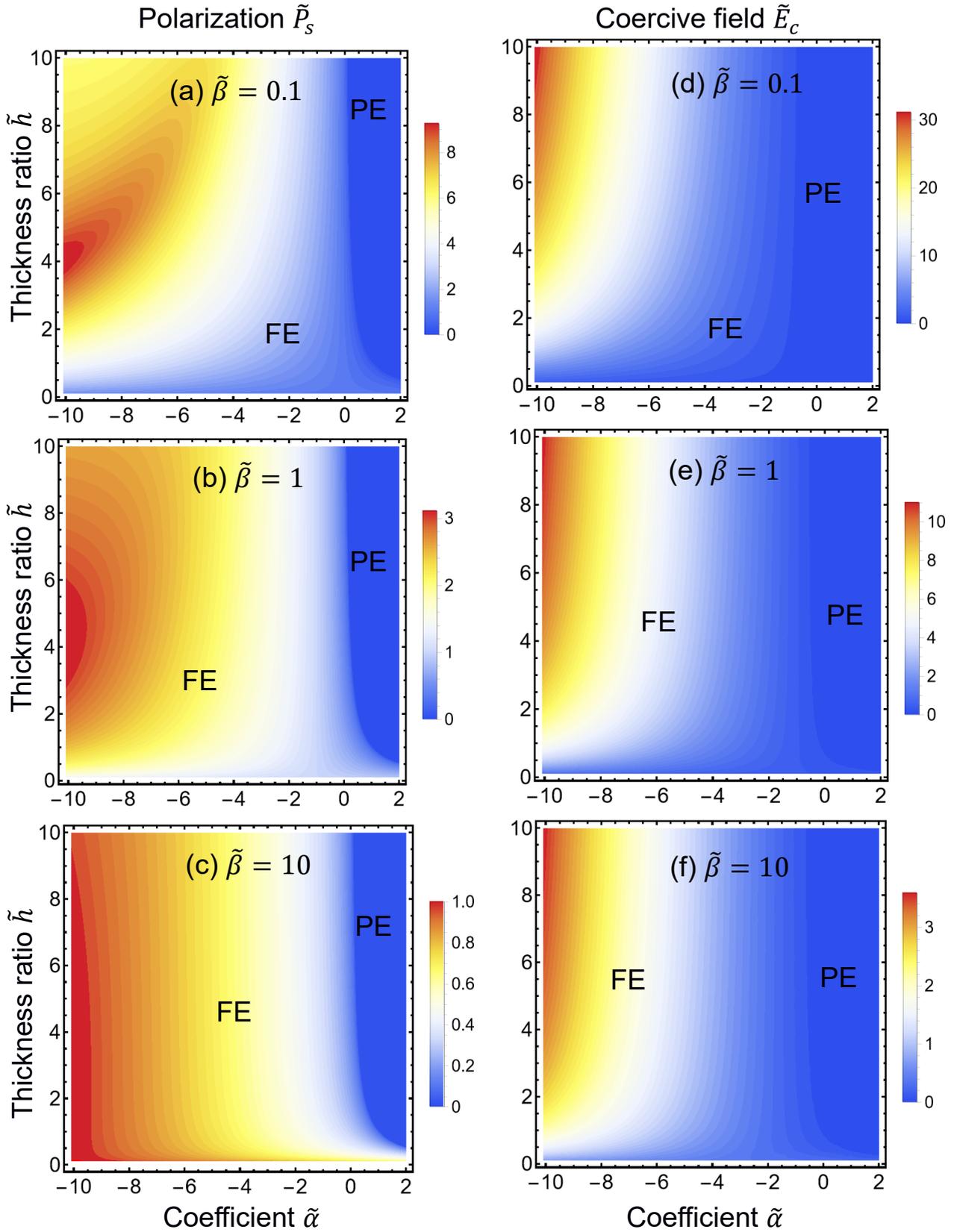

**FIGURE B2.** The dependences of the dimensionless remanent polarization $\tilde{P}_s$ **(a, b, c)** and coercive field $\tilde{E}_c$ **(d, e, f)** of the bilayer polarization hysteresis loop on the dimensionless LGD coefficient $\tilde{\alpha}$ and thickness ratio $\tilde{h}$ calculated for different values of dimensionless LGD coefficient $\tilde{\beta}$=0.1 **(a, d)**, 1 **(b, e)** and 10 **(c, f)**. Other parameters are the same as in **Fig. 2**.

**Appendix C. The estimation of LGD model parameters for the bulk $Al_{1-x}Sc_xN$ and AlN**

In the absence of internal electric field, the evident form of the LGD equations for the polarization $P_{si}$ of the layer "$i$" is

$$\alpha_i P_{si} + \beta_i P_{si}^3 + \gamma_i P_{si}^5 = 0. \qquad (C.1a)$$

Corresponding dielectric permittivity $\varepsilon_i$ is

$$\varepsilon_i = \varepsilon_{bi} + \frac{1}{\varepsilon_0(\alpha_i + 3\beta_i P_{si}^2 + 5\gamma_i P_{si}^4)}. \qquad (C.1b)$$

The available experimental results on the polarization and permittivity of the ferroelectric compounds $Al_{0.73}Sc_{0.27}N$ and AlN are summarized in **Table CI**.

**Table CI.** The parameters of $Al_{1-x}Sc_xN$ and AlN determined experimentally

| compound | $P_s$, µC/cm$^2$ | $\varepsilon$* | $\varepsilon_b$** | Note and reference |
|---|---|---|---|---|
| $Al_{0.73}Sc_{0.27}N$ | 110 | 17 | 4 | $P_s$ value taken from Ref. [43] |
| AlN | 127 | 10 | 4 | $P_s$ is estimated from experimental results of [44] |

*Dielectric permittivity $\varepsilon$ values taken from Fichtner et al. [45].

**Background permittivity $\varepsilon_b$ is estimated as the square of refractive index, which is about 2.1-2.3 for $Al_{1-x}Sc_xN$ and AlN according to Ref. [46].

In supposition that higher order coefficients are independent on the Sc-content, we could use the equations like (C.1) as a linear system for the unknown coefficients $\alpha_1$, $\alpha_2$, $\beta$ and $\gamma$, namely

$$\begin{cases} \alpha_1 + \beta P_{s1}^2 + \gamma P_{s1}^4 = 0, \\ \alpha_2 + \beta P_{s2}^2 + \gamma P_{s2}^4 = 0, \\ \alpha_1 + 3\beta P_{s1}^2 + 5\gamma P_{s1}^4 = \frac{1}{\varepsilon_0(\varepsilon_1 - \varepsilon_b)}, \\ \alpha_2 + 3\beta P_{s2}^2 + 5\gamma P_{s2}^4 = \frac{1}{\varepsilon_0(\varepsilon_2 - \varepsilon_b)}. \end{cases} \qquad (C.2)$$

Where $\alpha_1$ and $\alpha_2$ are the linear dielectric stiffness for the two compounds. The solution of (C.2) with the experimentally observed values of $P_{si}$ and $\varepsilon_i$ ($i$=1, 2) from **Table CI** is presented in **Table CII**.

**Table CII.** LGD model parameters of $Al_{1-x}Sc_xN$ and AlN

| compound | $\alpha$, m/F | $\beta$, m$^5$/(F C$^2$) | $\gamma$, m$^7$/(F C$^4$) | $g_z^{(i)}$, m$^3$/F | $g_\perp^{(i)}$, m$^3$/F |
|---|---|---|---|---|---|
| $Al_{0.73}Sc_{0.27}N$ | $-2.644 \cdot 10^8$ | $-3.155 \cdot 10^9$ | $2.788 \cdot 10^9$ | $5 \cdot 10^{-10}$ | $1 \cdot 10^{-10}$ |
| AlN | $-2.164 \cdot 10^9$ | $-3.155 \cdot 10^9$ | $2.788 \cdot 10^9$ | $5 \cdot 10^{-10}$ | $1 \cdot 10^{-10}$ |

Further we approximate the composition $Al_{1-x}Sc_xN$ dependence of $\alpha_i$ as follows

$$\alpha_i \cong \alpha(x) = \alpha_0 \left(1 - \frac{x}{x_c}\right) \qquad (C.3)$$

Where $x_c = 0.31$ is the sort of a critical concentration

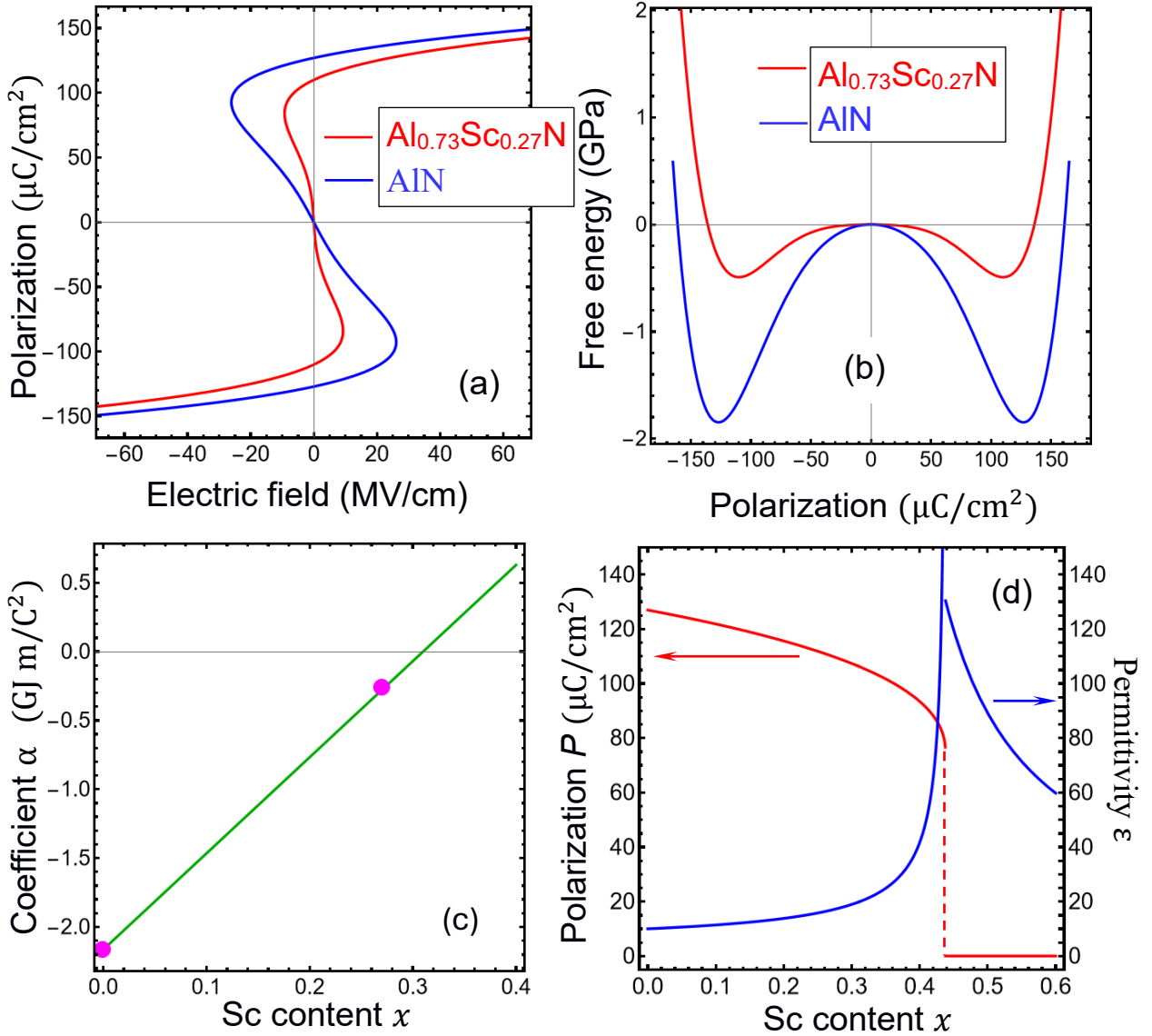

**FIGURE C1. (a)** Thermodynamic curves of polarization dependence on the electric field. **(b)** The free energy dependence on the polarization for $Al_{0.73}Sc_{0.27}N$ and AlN compounds (red and blue curves, respectively). Calculated composition dependence of the linear dielectric stiffness $\alpha(x)$ **(c)**, spontaneous polarization (red curve) and dielectric permittivity (blue curve) **(d)** of the $Al_{1-x}Sc_xN$.

## Appendix D. The estimation of LGD model parameters for the bulk $Zn_{1-x}Mg_xO$

LGD equation for the polarization $P_z$ of the uniaxial ferroelectric is

$$\alpha P_z + \beta P_z^3 + \gamma P_z^5 = E_z, \tag{D.1a}$$

where $E_z$ is the electric field component co-directed with the polar c-axis Z. Corresponding static dielectric permittivity $\varepsilon_{stat}$ is

$$\varepsilon_{stat} = \varepsilon_b + \frac{1}{\varepsilon_0(\alpha + 3\beta P_S^2 + 5\gamma P_S^4)}. \tag{D.1b}$$

The background contribution $\varepsilon_b$ to the dielectric susceptibility can be estimated as the square of the refraction index along the polar axis,

$$\varepsilon_b \cong n_c^2. \quad (D.2)$$

Schmidt et al. [47] found that the refraction index along the c-axis of wurtzite Zn$_{1-x}$Mg$_x$O depends on its composition $x$ as

$$n_c = 1.844 - 0.782x, \quad \text{for Zn}_{1-x}\text{Mg}_x\text{O with } 0 \leq x \leq 0.19. \quad (D.3)$$

The fitting parameters of the so-called dielectric function of a chemically pure cubic MgO could be used to estimate its dielectric permittivity as $\varepsilon_{stat} \approx 9.83$ and $\varepsilon_b \approx 3.02$. [48]

To find the parameters of LGD free energy expansion of Zn$_{1-x}$Mg$_x$O, we use the experimental results of Ferri et al. [49] who measured both the quasi-static dielectric constant at low field and the ferroelectric hysteresis loops in Zn$_{1-x}$Mg$_x$O with $x = 0.34$ and $0.37$. The spontaneous polarization of ZnO was estimated from the first principles calculations [50] being close to 0.9 C/m$^2$, and its dielectric permittivity value was taken from the experimental results of Hofmeister et al. [51]. The determined polar and dielectric properties of Zn$_{1-x}$Mg$_x$O ($0 \leq x \leq 1$) are summarized in **Table DI**.

**Table DI**. The static dielectric permittivity and polar properties of Zn$_{1-x}$Mg$_x$O

| x | 0 | 0.34 | 0.37 | 1 |
|---|---|---|---|---|
| $\varepsilon_{stat}$ | 11.3 | 18 | 16 | 9.83 |
| $P_s$, C/m$^2$ | 0.9 | 1.184 | 1.077 | 0 |
| $\varepsilon_b$ | 3.40 | 2.49 | 2.42 | 3.02 |

No hysteresis loops are observed experimentally in the bulk ZnO, and therefore it is considered as a non-switchable pyroelectric, whose hypothetic thermodynamic coercive field in higher than the electric breakdown field. Using use equations like (D.1) at $E_z = 0$ and parameters ($P_s$, $\varepsilon_b$ and $\varepsilon_{stat}$) from **Table DI** a linear system for the unknown coefficients $\alpha$, $\beta$ and $\gamma$ acquires the form

$$\begin{cases} \alpha P_s + \beta P_s^3 + \gamma P_s^5 = 0, \\ \alpha + 3\beta P_s^2 + 5\gamma P_s^4 = \dfrac{1}{\varepsilon_0(\varepsilon_{stat}-\varepsilon_b)}. \end{cases} \quad (D.4)$$

Using known values of spontaneous polarization $P_s$ and static dielectric permittivity $\varepsilon_{stat}$. However, the system (D.4) is two equations for the three independent coefficients $\alpha$, $\beta$ and $\gamma$, and thus the additional condition is needed. Due to the absence of ferroelectric hysteresis loops in ZnO, it is reasonable to assume that $\gamma = 0$. As for MgO, here $P_s = 0$ and only the coefficient $\alpha$ can be determined from Eqs.(D.4). To determine LGD coefficients $\alpha$, $\beta$ and $\gamma$ for Zn$_{1-x}$Mg$_x$O with $x = 0.34$ and $x = 0.37$ we used the ferroelectric hysteresis loops observed experimentally by Ferri et al.[49] and Eqs.(D.4) along with the total shape of the loops (see **Fig. D1**). Note that the single-domain LGD models overestimate strongly the coercive field of hysteresis loops, hence only the points near zero field were used for the fitting.

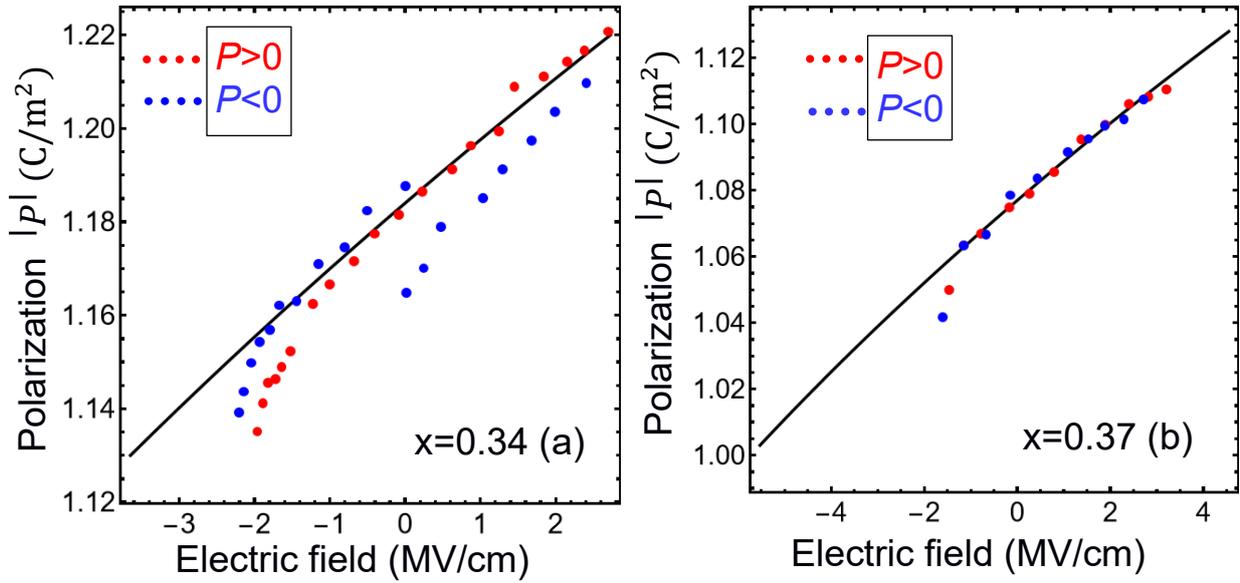

**Figure D1.** The polarization dependence on the electric field for $Zn_{1-x}Mg_xO$ with $x = 0.34$ **(a)** and 0.37 **(b)**. Black curves represent the fitting within the LGD model, Eqs.(D1a) and (D.4); red and blue circles are experimental results of Ferri et al. [49] for the upper and lower parts of the hysteresis loops, respectively. The fitting parameters are listed in **Table DII** along with the calculated values of the thermodynamic coercive field.

**Table DII.** LGD expansion coefficients for $Zn_{1-x}Mg_xO$ with $x = 0, 0.34, 0.37$ and $1$.

| x | $\varepsilon_b$ | $\alpha$, m/F | $\beta$, J m$^5$/C$^4$ | $\gamma$, J m$^9$/C$^6$ | $E_c^{th}$, MV/cm |
|---|---|---|---|---|---|
| 0 | 3.40 | $-7.152 \cdot 10^9$ | $8.829 \cdot 10^9$ | 0 | 24.7 |
| 0.34 | 2.49 | $2.854 \cdot 10^8$ | $-3.007 \cdot 10^9$ | $2.0 \cdot 10^9$ | 7.6 |
| 0.37 | 2.42 | $-3.870 \cdot 10^8$ | $-2.914 \cdot 10^9$ | $2.8 \cdot 10^9$ | 6.9 |
| 1 | 3.02 | $1.659 \cdot 10^{10}$ | 0 | 0 | 0 |